\documentclass[10pt,prd,singlecolumn,
nofootinbib,
noshowkeys,
noshowpacs,
superscriptaddress,
floatfix
]{revtex4-2}
\usepackage{amsmath,amsfonts,amsthm,amssymb}
\usepackage[dvips]{graphics,graphicx}
\usepackage[usenames,dvipsnames]{color}
\definecolor{darkblue}{RGB}{0,0,196}
\definecolor{darkgreen}{RGB}{0,120,0}
\usepackage[colorlinks=true,linktocpage=true,linkcolor=darkblue,citecolor=red,urlcolor=darkblue]{hyperref}
\usepackage{cancel}
\usepackage{bbold}
\usepackage{multirow}
\usepackage{longtable}
\usepackage{color}
\usepackage[normalem]{ulem}
\usepackage{hyperref}
\usepackage{bigints}
\usepackage{xparse}
\usepackage{physics}
\usepackage{verbatim}
\usepackage{minibox}
\usepackage{comment}
\usepackage{appendix}
\usepackage{slashed}
\usepackage{marginnote}
\usepackage{graphicx}
\usepackage[nice]{nicefrac}
\usepackage{amsmath}
\usepackage[colorinlistoftodos]{todonotes}
\usepackage{hepunits}
\usepackage{stackengine,scalerel}
\newcommand\hstar[1]{\ThisStyle{\ensurestackMath{%
  \setbox0=\hbox{$\SavedStyle#1$}%
  \stackengine{0pt}{\copy0}{\kern.2\ht0\smash{\SavedStyle\star}}{O}{c}{F}{T}{S}}}}

\definecolor {darkgreen}{rgb}{0.2,0.7,0.2}

\begin{document}
\title{Relativistic second-order spin hydrodynamics: an entropy-current analysis}

\author{Rajesh Biswas}
\email{rajesh.biswas@pwr.edu.pl}
\affiliation{Institute for Theoretical Physics, Wroc\l{}aw University of Science and Technology, PL-50-370 Wroc\l{}aw, Poland}

\author{Asaad Daher}
\email{asaad.daher@ifj.edu.pl}
\affiliation{Institute  of  Nuclear  Physics  Polish  Academy  of  Sciences,  PL-31-342  Krak\'ow,  Poland}

\author{Arpan Das}
\email{arpan.das@ifj.edu.pl}
\affiliation{Institute  of  Nuclear  Physics  Polish  Academy  of  Sciences,  PL-31-342  Krak\'ow,  Poland}

\author{Wojciech Florkowski}
\email{wojciech.florkowski@uj.edu.pl}
\affiliation{Institute of Theoretical Physics, Jagiellonian University, PL-30-348 Krak\'ow, Poland}

\author{Radoslaw Ryblewski}
\email{radoslaw.ryblewski@ifj.edu.pl}
\affiliation{Institute  of  Nuclear  Physics  Polish  Academy  of  Sciences,  PL-31-342  Krak\'ow,  Poland}

%%%%%%%%%%%%%%%%%

\begin{abstract}
We present a new derivation of Israel-Stewart-like relativistic second-order dissipative spin hydrodynamic equations using the entropy current approach. In our analysis, we consider a general energy-momentum tensor with symmetric and anti-symmetric parts. Moreover, the spin tensor, which is not separately conserved, has a simple phenomenological form that is antisymmetric only in the last two indices. Apart from the evolution equations for energy density, fluid flow, and spin density, we also find relaxation-type dynamical equations for various dissipative currents. The 
latter are consistently derived within the second-order theory as gradient corrections to the energy-momentum and spin tensors. We argue that this approach correctly reproduces the corresponding Navier-Stokes limit of spin hydrodynamic equations. Throughout our analysis, the spin chemical potential is considered a $\mathcal{O}(\partial)$ quantity in the hydrodynamic gradient expansion and reduces to thermal vorticity in the global equilibrium. New coefficients appearing in the generalized spin hydrodynamic equations are undetermined and can only be evaluated within a proper underlying microscopic theory of a given system.
\end{abstract}

%We derive the relativistic second-order spin hydrodynamic evolution equations using entropy-current analysis. Apart from the basic evolution of energy, linear momentum, and total angular momentum, we obtain the fluid-dynamical equations of the dissipative currents composing the symmetric and antisymmetric parts of the energy-momentum tensor. The novel feature of the second-order theory is fixing the evolution equation of the 3-rank spin tensor $S^{\lambda\mu\nu}$, which follows from the conservation of total angular momentum, by analyzing various evolution of its dissipative part $S^{\lambda\mu\nu}_{(1)}$.

\maketitle
%%%%%%%%%%%%%%%%%%%%%%%%%%%%%%%%%%%%%%%%%%%%%%
\section{Introduction}
In non-central relativistic heavy-ion collisions, the average spin polarization of hadrons (e.g., $\Lambda$ hyperons) is observed along the global axis of rotation of the produced matter~\cite{STAR:2017ckg, STAR:2018gyt,STAR:2019erd,ALICE:2019onw,ALICE:2019aid,STAR:2020xbm,Kornas:2020qzi,STAR:2021beb,ALICE:2021pzu}. This result may suggest that constituents' spin in the hyperons is coordinated in a specific direction, implying that the quark-gluon plasma (QGP) contains non-trivial vortical structures~\cite{Liang:2004ph,Betz:2007kg}, which in turn might be caused by the significant amount of orbital angular momentum produced in such collisions~\cite{lisa2021,Becattini:2007sr}. This phenomenon mimics the Barnett effect~\cite{barnett1915magnetization,RevModPhys.7.129} which displays the macroscopic effect of a quantum spin. Various theoretical approaches have been explored to model the vortical structure of a QCD plasma, e.g., hydrodynamic approach~\cite{Hattori:2019lfp,Fukushima:2020ucl,Li:2020eon,She:2021lhe,Daher:2022xon,Hongo:2021ona,Speranza:2021bxf,Wang:2021ngp,Gallegos:2021bzp}, relativistic kinetic theory~\cite{Florkowski:2017ruc,Florkowski:2017dyn,Hidaka:2017auj,Florkowski:2018myy,Weickgenannt:2019dks,Bhadury:2020puc,Weickgenannt:2020aaf,Shi:2020htn,Speranza:2020ilk,Bhadury:2020cop,Singh:2020rht,Bhadury:2021oat,Peng:2021ago,Sheng:2021kfc,Sheng:2022ssd,Hu:2021pwh,Hu:2022lpi,Fang:2022ttm,Wang:2022yli}, effective Lagrangian approach~\cite{Montenegro:2017rbu,Montenegro:2017lvf,Montenegro:2018bcf,Montenegro:2020paq}, quantum statistical density operators~\cite{Becattini:2007nd,Becattini:2009wh,Becattini:2012pp,Becattini:2012tc,Becattini:2018duy,Hu:2021lnx}, holography~\cite{Gallegos:2020otk,Garbiso:2020puw}, etc. Considering the triumphs of the relativistic dissipative hydrodynamic frameworks in relativistic heavy-ion phenomenology~\cite{Florkowski:2017olj,Elfner:2022iae,Heinz:2013th}, several extensions of relativistic hydrodynamics with spin degrees of freedom for the vortical fluids attracted a lot of attention. The spin hydrodynamic frameworks have a crucial role to play in understanding the collective spin dynamics of relativistic strongly interacting plasma because they may link quantum mechanical features of matter with hydrodynamics. 
%\smallskip

To model the collective spin dynamics in relativistic spin hydrodynamic frameworks, in addition to the usual hydrodynamic quantities, e.g., the energy-momentum tensor ($T^{\mu\nu}$), one also introduces the 3-rank spin tensor ($S^{\lambda\mu\nu}$)~\cite{Florkowski:2017ruc}. The additional equations of motion resulting from the conservation of the system's total angular momentum provide information about the dynamical evolution of the spin tensor. One of the fundamental conceptual difficulties in formulating a theory of relativistic dissipative spin hydrodynamics is the problem of ``pseudo-gauge transformations". Pseudo-gauge transformations imply that the forms of the energy-momentum tensor and spin tensor are not unique. In particular, for any energy-momentum tensor $T^{\mu\nu}$ satisfying the conservation equation, i.e., $\partial_{\mu}T^{\mu\nu}=0$, one can construct an equivalent energy-momentum tensor $T^{\prime \,\mu\nu}$ by adding the divergence of an antisymmetric tensor, namely $T^{\prime \,\mu\nu} = T^{\mu\nu} + \partial_\lambda \Phi^{\nu\mu \lambda}$~\cite{Chen:2018cts,HEHL197655,Speranza:2020ilk}. Note that if $\Phi^{\nu\mu \lambda}$ is antisymmetric in the last two indices then $T^{\prime \,\mu\nu}$ is also conserved. The same construction of the spin tensor can also be obtained without affecting the conservation of the total angular momentum. Different pseudo-gauge choices do not affect the conservation of total angular momentum or energy-momentum, nor do these transformations have any impact on the global charges (i.e., the global energy, linear momentum, and angular momentum). Various pseudo-gauge choices, e.g., the canonical, Belinfante-Rosenfeld (BR) ~\cite{BELINFANTE1939887,BELINFANTE1940449,Rosenfeld1940}, de Groot-van Leeuwen-van  Weert  (GLW)~\cite{DeGroot:1980dk}, Hilgevoord-Wouthuysen (HW) ~\cite{HILGEVOORD19631,HILGEVOORD19651002} forms and their implications on the spin hydrodynamic framework are intensely debated in recent literature~\cite{Becattini:2018duy,Speranza:2020ilk,Fukushima:2020ucl,Li:2020eon,Buzzegoli:2021wlg,Das:2021aar,Daher:2022xon}. 

Without going into a specific microscopic theory, a model-independent dissipative spin hydrodynamic framework can be obtained using thermodynamic consideration, which implies that for a dissipative system, entropy must be produced. This simple but rather powerful physical principle has been implemented very rigorously to obtain the Navier-Stokes-like theory of dissipative spin hydrodynamic framework~\cite{Hattori:2019lfp,Fukushima:2020ucl,Daher:2022xon}. In this framework, the energy-momentum tensor consists of symmetric as well as antisymmetric components. Moreover, following the earlier works of Weyssenhoff and Raabe~\cite{Weyssenhoff:1947iua}, one considers a simple \textit{phenomenological} form of the spin tensor, which is only antisymmetric in the last two indices $S^{\lambda\mu\nu}=u^{\lambda}S^{\mu\nu}$~\cite{Hattori:2019lfp, Fukushima:2020ucl,Daher:2022xon}. Here $u^{\mu}$ represents the time-like fluid flow four vector, and $S^{\mu\nu}$ represents the spin density in analogy with the number density. A linear stability analysis for this \textit{phenomenological} first-order spin hydrodynamic framework has been performed in Refs.~\cite{Daher:2022wzf, Sarwar:2022yzs}. These analyses show that in the fluid rest frame, the first-order spin hydrodynamic equations are generally unstable under linear perturbation~\cite{Daher:2022wzf}. This is a rather interesting result because the instability manifests itself even in the fluid rest frame, and the source of this instability is the spin equation of state that relates the spin density tensor ($S^{\mu\nu}$) to the spin chemical potential ($\omega^{\mu\nu}$). Strictly speaking, it has been argued that only the spin density perturbation components $\delta S^{0i}$ are responsible for the instabilities. Also an independent analysis of this framework, for a boost invariant system indicates unstable behavior in the evolution of the temperature ($T$) and the spin chemical potential ($\omega^{\mu\nu}$)~\cite{Biswas:2022bht}. These instabilities can be generic and the first-order (Navier-Stokes limit) spin-hydrodynamic framework can be highly pathological. Second-order dissipative hydrodynamic frameworks have been argued to be free of stability as well as causality issues~\cite{Hiscock:1983zz,Hiscock:1987zz,Kovtun:2019hdm,Bemfica:2019knx,Koide:2006ef,Abbasi:2022rum,Denicol:2008ha,Van:2007pw,Pu:2009fj}. We expect that such features will also remain intact for second-order spin hydrodynamic frameworks. Such observation motivates us to go beyond the first-order theory.
%\smallskip

In this paper, we construct a new second-order Israel-Stewart-like (IS-like) spin hydrodynamic framework using the entropy current analysis~\cite{Israel:1979wp,Montenegro:2016gjq,Heinz:2014zha,Brito:2020nou}.  Some efforts have been already made to derive the second-order spin hydrodynamic equations from an underlying microscopic theory~\cite{Weickgenannt:2022zxs, Weickgenannt:2022qvh} using spin-kinetic equations. Such a kinetic-theory approach explicitly uses spin-dependent collision terms and is based on the moment method of kinetic equation. In this article, we follow an alternative model-independent way based on the entropy current analysis to derive the second-order spin hydrodynamic equations~\cite{Israel:1979wp}. Various second-order hydrodynamic theories for `spin-less' fluid, e.g., the Muller-Israel-Stewart (MIS) approach~\cite{Israel:1979wp,Israel:1976tn,Muller:1967zza}, Denicol-Niemi-Molnar-Rischke (DNMR) approach~\cite{Denicol:2010xn,Denicol:2012cn}, Baier-Romatschke-Son-Starinets-Stephanov (BRSSS) approach~\cite{Baier:2007ix}, Chapman-Enskog approach~\cite{Jaiswal:2012qm,Jaiswal:2013npa,Jaiswal:2013fc} etc., have been routinely used to explain the heavy-ion collision data. Although different second-order hydrodynamic theories can have a similar structure, they are not exactly the same which is reflected in the hydrodynamic evolution, particularly where the gradients are large~\cite{Florkowski:2016kjj}. Such differences crucially affect their application to explain the heavy ion collision data. These differences may also become evident for second-order spin hydrodynamic frameworks. The present calculation can be considered as a complementary method to the kinetic theory approach to obtain spin hydrodynamic equations.

After this brief introduction, in Sec.~\ref{secII} we discuss the Navier-Stokes theory of dissipative spin hydrodynamics using the entropy current analysis. Once the  Navier-Stokes theory is defined we next move to the construction of the second-order Israel-Stewart theory of dissipative spin hydrodynamics in sec~\ref{secIII}. Finally, in Sec.~\ref{secV} we conclude our results with an outlook. 

In this manuscript, the symmetric and antisymmetric parts of a tensor $X^{\mu\nu}$ are denoted as $X^{\mu\nu}_{(s)}\equiv X^{(\mu\nu)}\equiv (X^{\mu\nu}+X^{\nu\mu})/2$ and $X^{\mu\nu}_{(a)}\equiv X^{[\mu\nu]}\equiv (X^{\mu\nu}-X^{\nu\mu})/2$, respectively. We use the metric tensor of the signature $g_{\mu\nu}= \hbox{diag}(+1, -1, -1, -1)$ and the totally antisymmetric Levi-Civita tensor with the sign convention $\epsilon^{0123} = -\epsilon_{0123} = 1$. The fluid four-velocity $u^{\mu}$ satisfies the normalization condition $u^{\mu}u_{\mu}= 1$. The projector orthogonal to $u^{\mu}$ is defined as $\Delta^{\mu\nu}\equiv g^{\mu\nu}-u^{\mu}u^{\nu}$; by definition $\Delta^{\mu\nu}u_{\mu}=0$. Projection orthogonal to $u^{\mu}$ of a four-vector $X^\mu$ is represented as $X^{\langle\mu\rangle}\equiv \Delta^{\mu\nu}X_{\nu}$. Traceless  and symmetric projection operator orthogonal to $u^{\mu}$ is denoted as $X^{\langle\mu\nu\rangle}\equiv \Delta^{\mu\nu}_{\alpha\beta}X^{\alpha\beta}\equiv \frac{1}{2}\left(\Delta^{\mu}_{~\alpha}\Delta^{\nu}_{~\beta}+\Delta^{\mu}_{~\beta}\Delta^{\nu}_{~\alpha}-\frac{2}{3}\Delta^{\mu\nu}\Delta_{\alpha\beta}\right)X^{\alpha\beta}$. Similarly, $X^{\langle[\mu\nu]\rangle}\equiv \Delta^{[\mu\nu]}_{[\alpha\beta]}X^{\alpha\beta}\equiv \frac{1}{2}\left(\Delta^{\mu}_{~\alpha}\Delta^{\nu}_{~\beta}-\Delta^{\mu}_{~\beta}\Delta^{\nu}_{~\alpha}\right)X^{\alpha\beta}$ denotes the antisymmetric projection operator orthogonal to $u^{\mu}$.  
The partial derivative operator can be decomposed into two parts, one along the flow direction and the other orthogonal to it, i.e., $\partial_{\mu}=u_{\mu}D+\nabla_{\mu}$. Here $D\equiv u^{\mu}\partial_{\mu}$ denotes the comoving derivative, and $\nabla_{\mu}\equiv\Delta_{\mu}^{~\alpha}\partial_{\alpha}$ is orthogonal to $u^{\mu}$, i.e., $u_{\mu}\nabla^{\mu}=0$. The expansion rate is defined as $\theta\equiv \partial_{\mu}u^{\mu}$.  

%%%%%%%%%%%%%%%%%%%%%%%%%%%%%%%%%%%%%%%%%%%%%%
\section{First-order relativistic dissipative spin hydrodynamics}
\label{secII}
\subsection{Macroscopic conservation laws}
Phenomenological derivation of hydrodynamics for a spin-polarized fluid is based on the conservation of energy-momentum tensor $T^{\mu\nu}$ and total angular momentum tensor $J^{\lambda\mu\nu}$\footnote{For simplicity, we assume that the system has no other conserved currents.}~\cite{Florkowski:2017dyn,Florkowski:2017ruc},
\begin{align}
    \partial_{\mu}T^{\mu\nu}=0,\label{eq1}\\[5pt]
    \partial_{\lambda}J^{\lambda\mu\nu}=2T_{(a)}^{\mu\nu}+\partial_{\lambda}S^{\lambda\mu\nu}=0.\label{eq2}
\end{align}
The total angular momentum tensor, $J^{\lambda\mu\nu}\!=\!L^{\lambda\mu\nu}+S^{\lambda\mu\nu}$, is the sum of the spin part, $S^{\lambda\mu\nu}$, and the orbital part, $L^{\lambda\mu\nu}=2\,x^{[\mu}T^{\lambda\nu]}$. In principle, $T^{\mu\nu}$, and $S^{\lambda\mu\nu}$ can be obtained from a more fundamental energy-momentum tensor operator and spin operator of the underlying quantum field theory system. Utilizing Noether's theorem from the perspective of the quantum field theory of Dirac fermions, the microscopic \textit{canonical} energy-momentum tensor is in general asymmetric, and the corresponding spin tensor is totally antisymmetric~\cite{Kleinert2016ParticlesAQ}. We expect that the symmetry properties of various microscopic currents will also be preserved at the macroscopic level. Due to the pseudo-gauge transformation $T^{\mu\nu}$, and $S^{\lambda\mu\nu}$ are not unique.
Using the arbitrariness in defining the energy-momentum tensor and the spin tensor, for \textit{phenomenological} studies, one often uses an asymmetric energy-momentum tensor and a spin tensor that is only antisymmetric in the last two indices~\cite{Weyssenhoff:1947iua}. The dissipative spin hydrodynamic framework with the \textit{phenomenological} form of the spin tensor has been discussed in Refs.~\cite{Hattori:2019lfp,Fukushima:2020ucl}. Moreover, it can be shown that the \textit{phenomenological} spin-hydrodynamic framework with the spin tensor which is antisymmetric only in the last two indices can be obtained from a properly defined \textit{canonical} spin-hydrodynamic framework with totally antisymmetric spin tensor using a proper pseudo-gauge transformation~\cite{Daher:2022xon}. In this work, we will first overview the first-order dissipative \textit{phenomenological} spin-hydrodynamic framework by considering the following forms of the energy-momentum tensor and spin tensor,

\begin{align}
&T^{\mu\nu}_{\rm }=T^{\mu\nu}_{\rm (0)}+T^{\mu\nu}_{{\rm }(1s)}+T^{\mu\nu}_{{\rm }(1a)}=T^{\mu\nu}_{\rm (0)}
+2h^{(\mu}u^{\nu)}+\tau^{\mu\nu}+2q^{[\mu}u^{\nu]}
+\phi^{\mu\nu}, \label{eq3}\\[5pt]
&S^{\lambda\mu\nu}_{\rm }=S^{\lambda\mu\nu}_{(0)}+S^{\lambda\mu\nu}_{(1)}=u^{\lambda} S^{\mu \nu}+S^{\lambda\mu\nu}_{(1)}. \label{eq4}
\end{align}
The leading order contribution $T_{(0)}^{\mu\nu}$ in Eq.~\eqref{eq3} has the form of the perfect fluid energy-momentum tensor,
\begin{align}
T^{\mu\nu}_{\rm (0)}=\varepsilon u^{\mu}u^{\nu}-p \Delta^{\mu\nu},\label{perfect fluid EMT}
\end{align}
where $\varepsilon$ is the energy density and $p$ is the equilibrium pressure. The most general expression of $T^{\mu\nu}$ can contain terms that are symmetric as well as antisymmetric under the $\mu\leftrightarrow\nu$ exchange. Therefore, we decompose the dissipative part of the energy-momentum tensor $T^{\mu\nu}_{\rm (1)}$ into a symmetric part $T^{\mu\nu}_{{\rm }(1s)}\equiv2h^{(\mu}u^{\nu)}+\tau^{\mu\nu}$ and an antisymmetric part $T^{\mu\nu}_{{\rm }(1a)}=2q^{[\mu}u^{\nu]}+\phi^{\mu\nu}$. The vector $h^{\mu}$ represents the heat flow, while $\tau^{\mu\nu}$ is the symmetric part of the dissipative correction such that $\tau^{\mu\nu}=\pi^{\mu\nu}+\Pi\,\Delta^{\mu\nu}$. The tensor $\pi^{\mu\nu}$ (the traceless part of $\tau^{\mu\nu}$) is the shear stress tensor and $\Pi$ is the bulk pressure. Analogously, $q^{\mu}$ and $\phi^{\mu\nu}$ are the antisymmetric dissipative corrections. These dissipative currents satisfy the following conditions: $h^{\mu}u_{\mu}=0$, $\tau^{\mu\nu}u_{\nu}=0$, $q^{\mu}u_{\mu}=0$, $\phi^{\mu\nu}u_{\nu}=0$,
$\tau^{\mu\nu}=\tau^{\nu\mu}$, and $\phi^{\mu\nu}=-\phi^{\nu\mu}$. According to the hydrodynamic gradient expansion $\varepsilon$, $p$, and $u^{\mu}$ scale as $\mathcal{O}(\partial^0)$ or $\mathcal{O}(1)$. But $h^{\mu}$, $q^{\mu}$, $\tau^{\mu\nu}$, and $\phi^{\mu\nu}$ scale as $\mathcal{O}(\partial)$. The tensor $S^{\mu\nu}=-S^{\nu\mu}$ in Eq.~(\ref{eq4}) can be interpreted as the spin density, $S^{\mu\nu}=u_{\lambda}S^{\lambda\mu\nu}$, in analogy to the number density~\cite{Hattori:2019lfp, Fukushima:2020ucl,Daher:2022xon}. Consequently, the spin density is a leading order term in the hydrodynamic gradient expansion, i.e., $S^{\mu\nu}\sim\mathcal{O}(1)$. The first-order dissipative correction $S^{\lambda\mu\nu}_{(1)}$ satisfies $u_{\lambda}S^{\lambda\mu\nu}_{(1)}=0$. Note that in general, $u_{\mu}S^{\mu\alpha\beta}_{(1)}\neq 0$, but due to the matching condition where $S^{\mu\nu}$ can be identified as the equilibrium spin density we consider $u_{\mu}S^{\mu\alpha\beta}_{(1)}= 0$. The same matching condition also identifies $\varepsilon$ as the equilibrium energy density, i.e., $ T^{\mu\nu}_{(1)}u_{\mu}u_{\nu}=0$. Using Eqs.~\eqref{eq3}, and \eqref{eq4} back into Eqs.~\eqref{eq1} and \eqref{eq2} we obtain spin hydrodynamic equations, 
\begin{align}
D\varepsilon+(\varepsilon+p)\theta &= -\partial \cdot h+h^{\nu}D u_{\nu}+\tau^{\mu \nu}  \partial_{\mu} u_{\nu}-\partial \cdot q
-q^{\nu}D u_{\nu}  +\phi^{\mu \nu}  \partial_{\mu} u_{\nu}
,\nonumber\\
&= 2 \,h^{\mu}Du_{\mu} -\nabla \cdot (q+h)+\tau^{\mu \nu}  \partial_{\mu} u_{\nu} +\phi^{\mu \nu}  \partial_{\mu} u_{\nu}
,\label{eq6}\\
(\varepsilon+p) Du^{\alpha} - \nabla^{\alpha}  p
&= -(h \cdot \partial) u^{\alpha}-h^{\alpha}\theta
-\Delta^{\alpha}_{~\nu}D h^{\nu}-\Delta^{\alpha}_{~\nu} \partial_{\mu} \tau^{\mu \nu}\nonumber\\
& ~~~~~-(q \cdot \partial) u^{\alpha}+q^{\alpha}\theta
+\Delta^{\alpha}_{~\nu}D q^{\nu}-\Delta^{\alpha}_{~\nu} \partial_{\mu} \phi^{\mu \nu}
 , 
\nonumber \\
&= 
-(q +h)\cdot \nabla u^{\alpha}+(q^{\alpha}-h^{\alpha})\theta
+\Delta^{\alpha}_{~\nu}D q^{\nu}-\Delta^{\alpha}_{~\nu}D h^{\nu}\nonumber\\
& ~~~~~-\Delta^{\alpha}_{~\nu} \partial_{\mu} \tau^{\mu \nu}-\Delta^{\alpha}_{~\nu} \partial_{\mu} \phi^{\mu \nu}
 , 
\label{eq7} \\
\partial_{\lambda}(u^{\lambda}S^{\mu\nu})+\partial_{\lambda}S^{\lambda\mu\nu}_{(1)} 
&= -2(q^{\mu}u^{\nu}-q^{\nu}u^{\mu}+\phi^{\mu\nu}). 
\label{eq8}
\end{align}
Due to the difficulty in specifying the flow velocity, frame choices are crucial in the setting of dissipative hydrodynamics\footnote{The energy-momentum tensor $T^{\mu\nu}$ can typically have 16 independent components in four dimensions. In dissipative hydrodynamics, these 16 components correspond to $\varepsilon, p, u^{\mu}, h^{\mu}, \pi^{\mu\nu}, \Pi, q^{\mu}$, and $\phi^{\mu\nu}$. Due to the equation of state, the variables $\varepsilon$ and $p$ together give only one unknown, while $u^{\mu}$, $h^{\mu}$ and $q^{\mu}$ have three independent degrees of freedom due to the conditions $u^{\mu}u_{\mu}=1$, $h^{\mu}u_{\mu}=0$ and $q^{\mu}u_{\mu}=0$. Both $\pi^{\mu\nu}$ and $\phi^{\mu\nu}$ are orthogonal to $u^{\mu}$. But $\pi^{\mu\nu}$ is symmetric and traceless. Hence, it has only five independent degrees of freedom. The tensor $\phi^{\mu\nu}$ is antisymmetric, hence it has three independent components. The bulk pressure $\Pi$ is just a scalar representing one degree of freedom. This counting summarizes to nineteen independent components in the $T^{\mu\nu}$ rather than sixteen. Therefore we have the freedom to eliminate three degrees of freedom. The so-called frame choice or the definition of $u^{\mu}$ reduces the number of independent components to sixteen.}. In standard hydrodynamics (spinless fluid) a natural hydrodynamic frame choice is the Landau frame, $T^{\mu\nu}u_{\nu}=\varepsilon u^{\mu}$ with only a symmetric energy-momentum tensor. This implies $h^{\mu}=0$. But in the spin hydrodynamic frameworks in general due to the presence of an antisymmetric component, one has two alternatives: (1) we can apply the Landau frame choice but only in the symmetric part of $T^{\mu\nu}$. This implies that $h^{\mu}=0$. (2) Instead of applying the Landau frame condition only to the symmetric part of the $T^{\mu\nu}$, we can also include the antisymmetric part. In that case, we obtain $h^{\mu}+q^{\mu}=0$. This immediately implies that we can have $h^{\mu}$ and $q^{\mu}$ nonvanishing but satisfying together the Landau condition. In this paper, we will keep the discussions general without imposing any specific frame condition, unless otherwise stated.  
%\medskip 
%%%%%%%%%%%%%%%%%%%
\subsection{Thermodynamic relations}
In the presence of dynamical spin degrees of freedom, the laws of thermodynamics can be generalized to~\cite{Hattori:2019lfp, Fukushima:2020ucl,Daher:2022xon},
\begin{align}    &\varepsilon+p=Ts+\omega_{\alpha\beta}S^{\alpha\beta},\nonumber\\    &d\varepsilon=Tds+\omega_{\alpha\beta}dS^{\alpha\beta},\nonumber\\
&dp=sdT+S^{\alpha\beta}d\omega_{\alpha\beta}.
\label{therodynamiceq}
\end{align}
Here, $T$ is the temperature, $s$ is the entropy density, and $\omega_{\alpha\beta}$ can be interpreted as the spin chemical potential conjugated to the spin density $S^{\alpha\beta}$ such that $S^{\alpha\beta}=\partial p/\partial \omega_{\alpha\beta}$ at a fixed temperature $T$. The spin chemical potential is defined as a hydrodynamic variable in analogy with the chemical potential and distinguishes spin hydrodynamic frameworks from the standard hydrodynamic theories.  However, there is a fundamental difference between these quantities. The chemical potential is only allowed in hydrodynamics if the corresponding current is conserved, e.g., baryon chemical potential in the presence of a conserved baryon current. But the presence of spin chemical potential does not necessarily imply the conservation of macroscopic spin current. In the language of the quantum statistical density operator framework~\cite{Becattini:2012tc}, in local thermal equilibrium, the spin chemical potential can only be considered as a Lagrange multiplier~\cite{Florkowski:2018ahw}. However, in global equilibrium, in the presence of an antisymmetric component of the energy-momentum tensor, the spin chemical potential can be shown to be related to the thermal vorticity, $\varpi_{\mu\nu}=-\frac{1}{2}(\partial_{\mu}\beta_{\nu}-\partial_{\nu}\beta_{\mu})$~\cite{Florkowski:2018ahw}. Here $\beta^{\mu}=\beta u^{\mu}$ and $\beta$ is the inverse temperature field.
%\medskip

Apart from the presence of spin chemical potential, the hydrodynamic gradient ordering of spin-related quantities appearing in Eq.~\eqref{therodynamiceq} has been discussed earlier. Fixing the hydrodynamic gradient ordering of $\omega^{\alpha\beta}$ is not straightforward. Since it is expected that in global equilibrium the spin chemical potential can be expressed in terms of thermal vorticity $\varpi_{\mu\nu}$, it is rather natural to consider $\omega^{\mu\nu}\sim \mathcal{O}(\partial)$. But such a conclusion is only applicable if the energy-momentum tensor is asymmetric~\cite{Florkowski:2018ahw}. This is a non-trivial aspect of the spin hydrodynamic framework as compared to the standard hydrodynamic frameworks for \textit{spinless} fluids. In standard hydrodynamics, the derivative correction terms vanish at global equilibrium. But all gradient terms do not vanish in global equilibrium if we consider the most generalized flow configuration, which is also true for spin-hydrodynamics. Using the framework of the quantum statistical density operator, it can be shown that the most general flow configuration in global equilibrium, must fulfill the following conditions~\cite{Florkowski:2018fap}, 
\begin{align}
\partial_{\mu}\beta_{\nu}+\partial_{\nu}\beta_{\mu}=0, 
\quad \beta_{\nu}=b_{\nu}+\varpi_{\nu\lambda}x^{\lambda}, 
\quad \varpi_{\mu\nu}=-\frac{1}{2}(\partial_{\mu}\beta_{\nu}-\partial_{\nu}\beta_{\mu})= \rm{constant}.
\label{equilibriumcondition}
\end{align}
Here $\beta^{\mu}=\beta u^{\mu}$, $\beta=1/T$, $b_{\nu}$ is a constant four vector. The 2-rank antisymmetric tensor $\varpi^{\mu\nu}$ is the thermal vorticity, and one can clearly observe that it scales as $\mathcal{O}(\partial)$ in the hydrodynamic gradient expansion.
Thus, a generic global equilibrium allows for $\mathcal{O}(\partial)$ terms in the flow configuration. Consequently, the gradient ordering of the spin chemical potential $\omega^{\mu\nu}$ is a contentious issue in the setting of spin hydrodynamics and has serious ramifications for the formulation of the spin hydrodynamic framework. A natural question could be raised here on how to connect $S^{\mu\nu}\sim\mathcal{O}(1)$ and $\omega_{\mu\nu}\sim\mathcal{O}(\partial)$ when their hydrodynamic gradient orders do not match. This was recently discussed in Ref.~\cite{Biswas:2022bht} as a new spin equation of state was constructed to match the gradient orders of $S^{\mu\nu}$ and $\omega^{\mu\nu}$ without any further assumptions. Nonetheless, one can also consider different hydrodynamic gradient ordering of spin chemical potential, particularly when the energy-momentum tensor is symmetric. A spin hydrodynamic framework was discussed in Ref.~\cite{She:2021lhe} where the spin chemical potential is considered the leading order ($\mathcal{O}(1)$) in gradient order expansion. In this paper, we will only consider the spin hydrodynamic framework with $\omega^{\mu\nu}\sim \mathcal{O}(\partial)$.
%%%%%%%%%%%%%%%%%%%%%%%%%
\subsection{Constitutive relations for dissipative currents in the Navier-Stokes limit}
\label{sub2.2}

We observe that while there are in total twenty two independent components of $T^{\mu\nu}$ and $S^{\mu\nu}$, Eqs.~\eqref{eq6}-\eqref{eq8} constitute only ten equations for the ten independent variables $T, u^{\mu}$, and $\omega^{\mu\nu}$. Note that, the hydrodynamic ordering of the term $\partial_{\lambda}S^{\lambda\mu\nu}_{(1)}$ in Eq.~\eqref{eq8} is higher than the rest of the terms. Therefore, for the first-order dissipative theory, we can neglect $S^{\lambda\mu\nu}_{(1)}$. However, to close Eqs.~\eqref{eq6}-\eqref{eq8}, we still have to provide additional equations of motion for different dissipative currents. This eventually reduces to finding constitutive relations satisfied by the tensors $h^{\mu}$, $q^{\mu}$, $\Pi$, $\pi^{\mu\nu}$, and $\phi^{\mu\nu}$ in terms of $T, u^{\mu},$ and $\omega^{\mu\nu}$. Such constitutive relations can be obtained using the condition that, for a dissipative system, the entropy is no longer a conserved quantity but rather will be produced \cite{Israel:1979wp,Hattori:2019lfp}. The mathematical form of the entropy current within the framework of dissipative fluid dynamics is, a priori, not known. As a result, it is not trivial to obtain its evolution equation. However, one can proceed by first constructing the definition of the entropy current in the absence of derivative correction terms, i.e.,

\begin{equation}
\label{eq:9}
s_{}^{\mu}=\beta_{\nu}T^{\mu\nu}_{\rm (0)}+\beta^{\mu}p-\beta^{\mu}\omega_{\alpha\beta}S^{\alpha\beta}.
\end{equation}
Note that such a definition of equilibrium entropy current correctly reproduces equilibrium thermodynamic relation~\eqref{therodynamiceq} if we identify $s^{\mu}\equiv su^{\mu}$, where $s$ is the equilibrium entropy density. For an interacting fluid, we can generalize the definition of entropy current given above to incorporate dissipative terms. The non-equilibrium entropy current ansatz up to first-order in hydrodynamic gradient expansion, i.e., in the Navier-Stokes (NS) limit can be written as,
\begin{align}
    s^{\mu}_{\rm NS}&=\beta_{\nu}T^{\mu\nu}_{\rm }+\beta^{\mu}p-\beta\omega_{\alpha\beta}S^{\mu\alpha\beta}\nonumber\\[5pt]  
    &=\beta_{\nu}T^{\mu\nu}_{\rm (0)}+\beta_{\nu}T^{\mu\nu}_{\rm (1)}+\beta^{\mu}p
    -\beta^{\mu}\omega_{\alpha\beta}S^{\alpha\beta}
-\beta\omega_{\alpha\beta}S^{\mu\alpha\beta}_{(1)}\nonumber\\
    &=s^{\mu}_{\rm }+\beta_{\nu}T^{\mu\nu}_{\rm (1)}+\mathcal{O}(\partial^{2})\label{eq10},
\end{align}
where we make use of the equilibrium entropy current $s_{}^{\mu}$ defined in Eq.~\eqref{eq:9}. By imposing the second law of thermodynamics, i.e., $\partial_{\mu} s^{\mu}_{\rm NS}\geq 0$, for Eq.~\eqref{eq10}, we can obtain the constitutive relations of the various dissipative currents~\cite{Hattori:2019lfp,Daher:2022xon},
\begin{align}
     \Pi&=\zeta\theta, \label{eq14}\\[5pt] h^{\mu}  &=-\kappa\left(Du^{\mu}-\beta\nabla^{\mu}T\right), \label{eq11}\\[5pt]
     q^{\mu}&=\lambda\left(Du^{\mu}+\beta\nabla^{\mu}T-4\omega^{\mu\nu}u_{\nu}\right),\label{eq12}\\[5pt]
     \pi^{\mu\nu}&=2\eta\sigma^{\mu\nu},\label{eq13} \\[5pt]
     \phi^{\mu\nu}&=\gamma\left(\Omega^{\mu\nu}+2\beta\omega^{\langle\mu\rangle\langle\nu\rangle}\right)%,\nonumber\\~~~~~&
=\widetilde{\gamma}\left(2\nabla^{[\mu}u^{\nu]}+4\omega^{\langle\mu\rangle\langle\nu\rangle}\right).
    \label{eq15}
\end{align}
Here, all transport coefficients are positive, i.e., $\kappa\geq 0$, $\lambda\geq 0$, $\eta\geq 0$, $\zeta\geq 0$, and $\gamma\geq 0$.
We define $\widetilde{\gamma}=\beta\gamma/2$, $\sigma^{\mu\nu}=\nabla^{(\mu}u^{\nu)}-\frac{1}{3}\theta\Delta^{\mu\nu}=\Delta_{\alpha \beta}^{\mu \nu} \nabla^\alpha u^\beta$, $\Omega^{\mu\nu}=\beta\nabla^{[\mu}u^{\nu]}=\Delta^{\mu}_{\alpha}\Delta^{\nu}_{\beta}\partial^{[\alpha}\beta^{\beta]}$, and $\omega^{\langle\mu\rangle\langle\nu\rangle}=\Delta^{\mu\alpha}\Delta^{\nu\beta}\omega_{\alpha\beta}$. In these equations, all the terms on the r.h.s. are of order $\mathcal{O}(\partial)$ in hydrodynamic gradient expansion. Equations~\eqref{eq11}-\eqref{eq15} show explicitly that at this level, the number of state variables  $T, u^{\mu}, \omega^{\mu\nu}$ perfectly matches the number of dynamical equations~\eqref{eq6}-\eqref{eq8}. Note that if $\lambda=0$, and $\gamma=0$, then all the dissipative currents associated with the antisymmetric part of the energy-momentum tensor vanish. In this limit, if we consider the Landau frame choice, i.e., $h^{\mu}=0$, then nonvanishing dissipative currents are $\pi^{\mu\nu}$, and $\Pi$. Moreover, if we set $\omega^{\mu\nu}=0$, then the spin tensor also decouples from the theory. This is the NS limit giving rise to the standard hydrodynamics of \textit{spinless} fluid. Unfortunately, this first-order spin hydrodynamic framework can be shown to be pathological as it can give rise to instabilities under linear perturbations~\cite{Daher:2022wzf,Sarwar:2022yzs}. This is not a desired feature for a hydrodynamic theory, particularly for phenomenological applications.
%%%%%%%%%%%%%%%%%%%%%%%%%%%
\section{Towards second-order spin hydrodynamics}
\label{secIII}
\subsection{Entropy current for the second-order theory}
Historically, it is also well known that even for the spinless fluid, the relativistic NS theory is ill-defined because it can contain instabilities when perturbed around an arbitrary global equilibrium. The relativistic NS theory is unstable in the sense that small departures from equilibrium at one instant of time will diverge exponentially with time. The time scale of these instabilities can be short, which may affect the time evolution of the system~\cite{Hiscock:1985zz,Hiscock:1987zz}. We emphasize that in the comoving frame or in the rest frame, Landau's theory of dissipative hydrodynamics (for spinless fluid) is stable. However, the generic instability manifests itself in a Lorentz-boosted frame. Subsequently, it has been argued that such instabilities are intrinsically related to the acausal nature of the NS theory~\cite{Pu:2009fj}. Since the NS equations are not intrinsically hyperbolic, they allow for perturbations that propagate at an infinite speed. These fundamental problems provide overwhelming motivation to prohibit the practical application of relativistic NS theory. To incorporate dissipative effects consistently in fluid dynamics without violating causality, second-order theories are constructed, e.g., Israel-Stewart (IS) theory, etc. The IS second-order theory contains new parameters compared to the NS theory. Kinetic theory calculations have been used to show that these new parameters are nonvanishing and if these parameters are chosen appropriately then the dynamical equations governing the evolution of linear perturbations form a hyperbolic system of equations. Second-order dissipative hydrodynamic frameworks for spinless fluid have been argued to be free of stability and causality issues~\cite{Hiscock:1983zz,Hiscock:1987zz,Kovtun:2019hdm,Bemfica:2019knx,Koide:2006ef,Abbasi:2022rum,Denicol:2008ha,Van:2007pw,Pu:2009fj} which makes IS theory more acceptable as a viable hydrodynamic theory. We expect that such features will also remain intact for second-order spin hydrodynamic frameworks~\footnote{In the present calculation we develop the second-order theory for spin-hydrodynamics. Its stability and causality properties require extensive investigation which we will address in future works.}. Similarly to the NS theory here we also follow the entropy current analysis to derive the second-order spin hydrodynamic equations. In this approach once again we start with the entropy current for an arbitrary nonequilibrium state near equilibrium~\cite{Israel:1979wp},  
\begin{align}
    s^{\mu}_{\rm IS}&=\beta_{\nu}T^{\mu\nu}_{}+\beta^{\mu}p-\beta\omega_{\alpha\beta}S^{\mu\alpha\beta}+Q^{\mu},\nonumber\\
    &=\beta_{\nu}T^{\mu\nu}_{(0)}+\beta^{\mu}p-\beta^{\mu}\omega_{\alpha\beta}S^{\alpha\beta}+\beta_{\nu}T^{\mu\nu}_{\rm(1)}-\beta\omega_{\alpha\beta}S^{\mu\alpha\beta}_{(1)}+Q^{\mu},\nonumber\\
    &=s^{\mu}_{\rm NS}-\beta\omega_{\alpha\beta}S_{(1)}^{\mu\alpha\beta}+Q^{\mu}.\label{eq16}
\end{align}
Here $s^{\mu}_{\rm NS}$ contains the first-order corrections ($\mathcal{O}(\partial)$). The term $\beta\omega_{\alpha\beta}S_{(1)}^{\mu\alpha\beta}$ is second-order ($\mathcal{O}(\partial^2)$) in the hydrodynamic gradient expansion. Such a term does not appear in the NS limit, see Eq.~\eqref{eq10}. Novel information about new spin dissipative currents is embedded in $S^{\lambda\mu\nu}_{(1)}$ (Eq.~\eqref{eq4}). The term $Q^{\mu}$ is a general four vector containing terms up to second order ($\mathcal{O}(\partial^2))$. However, the form of $Q^{\mu}$ is not completely arbitrary as it contains all second-order terms composed of $h^{\mu}$, $\pi^{\mu\nu}$, $\Pi$, $q^{\mu}$, $\phi^{\mu\nu}$, and $S^{\mu\alpha\beta}_{(1)}$. The form of $Q^{\mu}$ is constrained by the condition that entropy is maximum in the equilibrium state. Contracting Eq.~\eqref{eq16} with $u^{\mu}$ we immediately obtain, $s_{\rm IS}-s=u_{\mu}Q^{\mu}$, where $s_{\rm IS}\equiv u_{\mu}s^{\mu}_{\rm IS}$. The condition that $s_{\rm IS}\leq s$ implies $u_{\mu}Q^{\mu}\leq 0$ (see Appendix~\ref{app1} for details). Before we introduce the most general expression of $Q^{\mu}$ we first express $S^{\mu\alpha\beta}_{(1)}$ in terms of irreducible tensors. Recall that the first-order correction to the spin tensor satisfies $u_{\mu}S^{\mu\alpha\beta}_{(1)}=0$ and it is antisymmetric in the last two indices. Therefore, the most general decomposition of $S^{\mu\alpha\beta}_{(1)}$ in terms of irreducible tensors takes the form~\cite{Becattini:2011ev} (see Appendix~\ref{appendixH}),  
\begin{align}
S^{\mu\alpha\beta}_{(1)}=2u^{[\alpha}\Delta^{\mu\beta]}\Phi+2u^{[\alpha}\tau^{\mu\beta]}_{(s)}+2u^{[\alpha}\tau^{\mu\beta]}_{(a)}+\Theta^{\mu\alpha\beta}.\label{eq17}
\end{align}
The new dissipative currents related to spin $\Phi,\tau^{\mu\nu}_{(s)}, \tau^{\mu\nu}_{(a)} $, and $\Theta^{\mu\alpha\beta}$ are of first-order in derivative expansion~$\mathcal{O}(\partial)$.
The currents satisfy the following properties: $u_{\mu} \tau_{(s)}^{\mu \beta}=$ $u_{\mu} \tau_{(a)}^{\mu \beta}=u_{\mu} \Theta^{\mu \alpha \beta}=0; \tau_{(s)}^{\mu \beta}=\tau_{(s)}^{\beta \mu}, \tau_{(a)}^{\mu \beta}=$ $-\tau_{(a)}^{\beta \mu}$, $\tau_{(s)\mu}^{~~\mu}=0$, $\Theta^{\mu \alpha \beta}=-\Theta^{\mu \beta \alpha}$, $u_{\mu}\Theta^{\mu \alpha \beta}=0$, $u_{\alpha}\Theta^{\mu \alpha \beta}=0$, and $u_{\beta}\Theta^{\mu \alpha \beta}=0$. Now we can express $Q^{\mu}$ in terms of all possible second-order combinations of dissipative currents respecting the constraint $u\cdot Q\leq 0$,
\begin{align}
Q^{\mu} = & ~~u^{\mu}\left(a_{1}\Pi^2+a_{2}\pi^{\lambda\nu}\pi_{\lambda\nu}+a_{3}h^{\lambda}h_{\lambda}+a_{4}q^{\lambda}q_{\lambda}+a_{5}\phi^{\lambda\nu}\phi_{\lambda\nu}\right)\nonumber
\\
\nonumber
& + u^{\mu}\left(\tilde{a}_{1}\Phi^{2}+\Tilde{a}_{2}\tau_{(s)}^{\lambda\nu}\tau_{(s)\lambda\nu}+\Tilde{a}_{3}\tau_{(a)}^{\lambda\nu}\tau_{(a)\lambda\nu}+\Tilde{a}_{4}\Theta^{\lambda\alpha\beta}\Theta_{\lambda\alpha\beta}\right)\nonumber
\\
& + \Big( b_{1}\Pi h^{\mu} + b_{2} \pi^{\mu\nu}h_{\nu}+b_{3}\phi^{\mu\nu}h_{\nu}+b_{4}\Pi q^{\mu}+ b_{5} \pi^{\mu\nu}q_{\nu}+b_{6}\phi^{\mu\nu}q_{\nu}\Big)\nonumber
\\
& + \left(\tilde{b}_{1}\Phi h^{\mu} + \tilde{b}_{2} \tau^{\mu\nu}_{(s)}h_{\nu}+\tilde{b}_{3}\tau^{\mu\nu}_{(a)}h_{\nu}+\tilde{b}_{4}\Phi q^{\mu}+ \tilde{b}_{5} \tau^{\mu\nu}_{(s)}q_{\nu}+\tilde{b}_{6}\tau^{\mu\nu}_{(a)}q_{\nu} \right)\nonumber
\\
&+\left(c_{1}\Theta^{\mu\alpha\beta}\phi_{\alpha\beta}+c_{2}\Theta^{\mu\alpha\beta}\tau_{(a)\alpha\beta}\right)\nonumber\\
& +\left(c_3 \Theta^{\alpha\beta\mu}\Delta_{\alpha\beta}\Pi+c_4 \Theta^{\alpha\beta\mu}\pi_{\alpha\beta}+c_5 \Theta^{\alpha\beta\mu}\Delta_{\alpha\beta}\Phi+c_6\Theta^{\alpha\beta\mu}\tau_{(s)\alpha\beta}\right)\nonumber\\
& +\left(c_7 \Theta^{\alpha\beta\mu}\phi_{\alpha\beta}+c_8\Theta^{\alpha\beta\mu}\tau_{(a)\alpha\beta}\right).
\label{eq19}
\end{align}
We define $a_{i}, \tilde{a}_{i}, b_{i}, \Tilde{b}_{i}, $ and $c_{i}$ to be dimensionful coefficients. While it is clear that due to $u\cdot Q\leq 0$ the $a (\tilde{a})$ coefficients have definite signatures with $a_1\leq 0$, $a_2\leq 0$, $a_3\geq 0$, $a_4\geq 0$, $a_5\leq 0$, $\tilde{a}_1\leq 0$, $\tilde{a}_2\leq 0$, $\tilde{a}_3\leq 0$, $\tilde{a}_4\geq 0$, there are no such sign constraints on $b_{i}, \Tilde{b}_{i}$, or $c_{i}$. Although a kinetic theory approach may indicate the sign of these coefficients. \\
%%%%%%%%%%%%%%%%%%%
\subsection{Evolution equations}
We argued that for the NS theory the dissipative currents $h^{\mu}$, $q^{\mu}$, $\pi^{\mu\nu}$, $\Pi$, and $\phi^{\mu\nu}$ can be expressed in terms of fundamental hydrodynamic variables $T, u^{\mu}$, and $\omega^{\mu\nu}$. This conclusion is obtained using the condition $\partial_{\mu}s^{\mu}_{\rm NS}\geq0$. But for the second-order theory, various dissipative currents are considered independent variables. This is evident from the fact that we have constructed second-order terms in $s^{\mu}_{\rm IS}$ in terms of these dissipative currents. Therefore, to close the hydrodynamic equations, we also need the evolution equation for these dissipative currents, which can be obtained using the condition that $\partial_{\mu}s_{\rm IS}^{\mu}\geq0$. Taking the divergence of $s^{\mu}_{\rm IS}$ and using spin-hydrodynamic equations, it can be shown that (see Appendix~\ref{appendixB} for details),  
\begin{align}
    \partial_{\mu}s^{\mu}_{\rm IS}=T^{\mu\nu}_{(1a)}\left(\partial_{\mu}\beta_{\nu}+2\beta\omega_{\mu\nu}\right)+\partial_{\mu}\beta_{\nu}T^{\mu\nu}_{(1s)}-\partial_{\mu}\left(\beta\omega_{\alpha\beta}\right)S^{\mu\alpha\beta}_{(1)}+\partial_{\mu}Q^{\mu}.\label{eq24}
\end{align}
Notice that for the global equilibrium condition $S^{\mu\alpha\beta}_{(1)}=0$, $Q^{\mu}=0$. Moreover, $\partial_{\mu}s^{\mu}_{\rm IS}=0$ implies the most general global equilibrium conditions~\eqref{equilibriumcondition}, i.e., the spin chemical potential converges to thermal vorticity, i.e., $\omega_{\mu\nu} \rightarrow
\frac{T}{2} \varpi_{\mu\nu}$ with $\beta_\mu=u_\mu/T$ satisfying the Killing condition $\partial_{(\mu}\beta_{\nu)}=0$. Using the explicit expressions for $T^{\mu\nu}_{(1s)}$, $T^{\mu\nu}_{(1a)}$ and $S^{\mu\alpha\beta}_{(1)}$, Eq.~\eqref{eq24} can be written as (see Appendix~\ref{appendixC} for details), 
\begin{align}
\partial_{\mu}s^{\mu}_{\rm IS}= &-\beta h^{\mu}\left(\beta \nabla_{\mu}T-Du_{\mu}\right)+\beta\pi^{\mu\nu}\sigma_{\mu\nu}+\beta\Pi \theta \nonumber\\
&-\beta q^{\mu}\left(\beta \nabla_{\mu}T+Du_{\mu}-4 \omega_{\mu\nu}u^{\nu}\right) +\phi^{\mu\nu}\left(\Omega_{\mu\nu}+2\beta \omega^{\langle\mu\rangle\langle\nu\rangle}\right)\nonumber\\
& -2\Phi u^{\alpha}\nabla^{\beta}(\beta\omega_{\alpha\beta})-2\tau^{\mu\beta}_{(s)}u^{\alpha}\Delta^{\gamma\rho}_{\mu\beta}\nabla_{\gamma}(\beta\omega_{\alpha\rho})-2\tau^{\mu\beta}_{(a)}u^{\alpha}\Delta^{[\gamma\rho]}_{[\mu\beta]}\nabla_{\gamma}(\beta\omega_{\alpha\rho})\nonumber\\
& -\Theta_{\mu\alpha\beta}\Delta^{\alpha\delta}\Delta^{\beta\rho}\Delta^{\mu\gamma}\nabla_{\gamma}(\beta\omega_{\delta\rho})+\partial_{\mu}Q^{\mu}. 
\label{equ22ver2}
\end{align}

As a last step, we need to investigate the term $\partial_{\mu}Q^{\mu}$ which can be done using the expression of $Q^{\mu}$ given in Eq.~\eqref{eq19}. A straightforward calculation gives, 
\begin{align}
   \partial_{\mu}Q^{\mu}&=h_{\alpha}\mathcal{A}^{\alpha}+q_{\alpha}\mathcal{B}^{\alpha}+\pi_{\alpha\beta}\mathcal{C^{\alpha\beta}}+\Pi \mathcal{D}+\phi_{\alpha\beta}\mathcal{E^{\alpha\beta}}\nonumber\\
   & +\Phi \mathcal{F}+\tau^{\alpha\beta}_{(s)}\mathcal{G_{\alpha\beta}}+\tau^{\alpha\beta}_{(a)}\mathcal{H_{\alpha\beta}}+\Theta_{\alpha\beta\gamma}\mathcal{I}^{\alpha\beta\gamma}.\label{equ23ver2}
\end{align}
In the above equations, scalars $\mathcal{D}$ and $\mathcal{F}$,   vectors $\mathcal{A}_{\beta}$ and  $\mathcal{B}_{\beta}$, and tensors $\mathcal{C}_{\mu\nu}$, $\mathcal{E_{\mu\nu}}$, $\mathcal{G_{\mu\nu}}$, $\mathcal{H}_{\mu\nu}$, and $\mathcal{I}_{\mu\nu\delta}$ are defined in Appendix~\ref{appendixD}.
Note that the dissipative fluxes multiplying these quantities satisfy the following properties: $h^{\mu}$ and $q^{\mu}$ are orthogonal to $u^{\mu}$, $\pi^{\mu\nu}$ and $\tau^{\mu\nu}_{(s)}$ are also orthogonal to $u^{\mu}$ as well as symmetric and traceless, $\phi^{\mu\nu}$ and $\tau^{\mu\nu}_{(a)}$ are orthogonal to $u^{\mu}$ as well as antisymmetric,  $\Theta^{\mu\alpha\beta}$ is antisymmetric in the last two indices and orthogonal to the fluid flow in all the indices. Using these properties
Eq.~\eqref{equ23ver2} can be expressed as,  

\begin{align}
   \partial_{\mu}Q^{\mu}&=h_{\alpha}\mathcal{A}^{\langle\alpha\rangle}+q_{\alpha}\mathcal{B}^{\langle\alpha\rangle}+\pi_{\alpha\beta}\mathcal{C^{\langle\alpha\beta\rangle}}+\Pi \mathcal{D}+\phi_{\alpha\beta}\mathcal{E^{\langle[\alpha\beta]\rangle}}\nonumber\\
   & +\Phi \mathcal{F}+\tau^{\alpha\beta}_{(s)}\mathcal{G_{\langle\alpha\beta\rangle}}+\tau^{\alpha\beta}_{(a)}\mathcal{H_{\langle[\alpha\beta]\rangle}}+\Theta_{\alpha\beta\gamma}\mathcal{I}^{\langle\alpha\rangle\langle\beta\rangle\langle\gamma\rangle}.\label{eq33}
\end{align}
The quantities 
$\mathcal{A}^{\langle\alpha\rangle}$, $\mathcal{B}^{\langle\alpha\rangle}$, $\mathcal{C^{\langle\alpha\beta\rangle}}$, $\mathcal{E^{\langle[\alpha\beta]\rangle}}$, $\mathcal{G^{\langle\alpha\beta\rangle}}$, $\mathcal{H^{\langle[\alpha\beta]\rangle}}$, and $\mathcal{I}^{\langle\alpha\rangle\langle\beta\rangle\langle\gamma\rangle}$ satisfy the following constraints,
\begin{align}
    & \mathcal{A}^{\langle\alpha\rangle}\equiv\Delta^{\alpha\beta}\mathcal{A}_{\beta}; \quad u_{\alpha}\mathcal{A}^{\langle\alpha\rangle}=0,\label{A}\\
    & \mathcal{B}^{\langle\alpha\rangle}\equiv\Delta^{\alpha\beta}\mathcal{B}_{\beta}; \quad u_{\alpha}\mathcal{B}^{\langle\alpha\rangle}=0,\label{B}\\
    & \mathcal{C}_{\langle\alpha\beta\rangle}\equiv\Delta_{\alpha\beta}^{\mu\nu}\mathcal{C}_{\mu\nu}=\frac{1}{2}\left(\Delta^{\mu}_{~\alpha}\Delta^{\nu}_{~\beta}+\Delta^{\mu}_{~\beta}\Delta^{\nu}_{~\alpha}-\frac{2}{3}\Delta_{\alpha\beta}\Delta^{\mu\nu}\right)\mathcal{C}_{\mu\nu}; \quad u^{\alpha}\mathcal{C}_{\langle\alpha\beta\rangle}=0; \quad g^{\alpha\beta}\mathcal{C}_{\langle\alpha\beta\rangle}=0,\label{C} \\
    & \mathcal{E}_{\langle[\alpha\beta]\rangle}\equiv \Delta^{[\mu\nu]}_{[\alpha\beta]}\mathcal{E}_{\mu\nu}\equiv\frac{1}{2}\left(\Delta^{\mu}_{~\alpha}\Delta^{\nu}_{~\beta}-\Delta^{\nu}_{~\alpha}\Delta^{\mu}_{~\beta}\right)\mathcal{E}_{\mu\nu}; \quad u_{\alpha}\mathcal{E^{\langle[\alpha\beta]\rangle}}=0, \label{E}\\
    & \mathcal{G_{\langle\alpha\beta\rangle}}\equiv\Delta^{\mu\nu}_{\alpha\beta}\mathcal{G_{\mu\nu}}; \quad u_{\alpha}\mathcal{G^{\langle\alpha\beta\rangle}}=0, \quad g_{\alpha\beta}\mathcal{G^{\langle\alpha\beta\rangle}}=0,\label{G}\\
    & \mathcal{H}_{\langle[\alpha\beta]\rangle}\equiv \Delta^{[\mu\nu]}_{[\alpha\beta]}\mathcal{H}_{\mu\nu}\equiv\frac{1}{2}\left(\Delta^{\mu}_{~\alpha}\Delta^{\nu}_{~\beta}-\Delta^{\nu}_{~\alpha}\Delta^{\mu}_{~\beta}\right)\mathcal{H}_{\mu\nu}; \quad u_{\alpha}\mathcal{H^{\langle[\alpha\beta]\rangle}}=0,\label{H}\\
    & \mathcal{I}^{\langle\alpha\rangle\langle\beta\rangle\langle\gamma\rangle}\equiv\Delta^{\alpha\mu}\Delta^{\beta\nu}\Delta^{\gamma\delta}\mathcal{I}_{\mu\nu\delta};\quad u_{\alpha}\mathcal{I}^{\langle\alpha\rangle\langle\beta\rangle\langle\gamma\rangle}=0; \quad u_{\beta}\mathcal{I}^{\langle\alpha\rangle\langle\beta\rangle\langle\gamma\rangle}=0;\quad u_{\gamma}\mathcal{I}^{\langle\alpha\rangle\langle\beta\rangle\langle\gamma\rangle}=0.\label{I2}
    \end{align}
Using Eq.~\eqref{eq33} in Eq.~\eqref{equ22ver2} the full form of the divergence of entropy current in the second-order theory can be written as
\begin{align}
   \partial_{\mu}s^{\mu}_{\rm IS}=&-\beta h^{\mu}\left(\beta\nabla_{\mu}T-Du_{\mu}-T\mathcal{A}_{\langle\mu\rangle}\right)+\beta\pi^{\mu\nu}\left(\sigma_{\mu\nu}+T\mathcal{C}_{\langle\mu\nu\rangle}\right)+\beta\Pi\left(\theta+T\mathcal{D}\right)\nonumber\\
   &-\beta q^{\mu}\left(\beta\nabla_{\mu}T+Du_{\mu}-4\omega_{\mu\nu}u^{\nu}-T\mathcal{B}_{\langle\mu\rangle}\right)   
+\phi^{\mu\nu}\left(\Omega_{\mu\nu}+2\beta\omega_{\langle\mu\rangle\langle\nu\rangle}+\mathcal{E}_{\langle[\mu\nu]\rangle}\right)\nonumber\\
   &
   +\Phi\left[-2 u^{\alpha}\nabla^{\beta}(\beta\omega_{\alpha\beta})+\mathcal{F}\right]
   +\tau^{\mu\beta}_{(s)}\left[-2u^{\alpha}\Delta^{\gamma\rho}_{\mu\beta}\nabla_{\gamma}(\beta\omega_{\alpha\rho})+\mathcal{G}_{\langle\mu\beta\rangle}\right]\nonumber\\
   &
   +\tau^{\mu\beta}_{(a)}\left[-2u^{\alpha}\Delta^{[\gamma\rho]}_{[\mu\beta]}\nabla_{\gamma}(\beta\omega_{\alpha\rho})
   +\mathcal{H}_{\langle[\mu\beta]\rangle}\right]
   +\Theta^{\mu\alpha\beta}\left[-\Delta^{\delta}_{~\alpha}\Delta^{\rho}_{~\beta}\Delta^{\gamma}_{~\mu}\nabla_{\gamma}(\beta\omega_{\delta\rho})+\mathcal{I}_{\langle\mu\rangle\langle\alpha\rangle\langle\beta\rangle}\right]
   \label{equ32ver2}
\end{align} 
Similarly to the NS theory the condition $\partial_{\mu}s^{\mu}_{\rm IS}\geq0$ gives us the following relations involving various dissipative currents appearing in the energy-momentum tensor,
\begin{align}
& \Pi=\zeta\big(\theta+T\mathcal{D}\big)\label{equ33ver2}\\
& h^{\mu}=-\kappa\left(Du^{\mu}-\beta\nabla^{\mu}T+T\mathcal{A^{\langle\mu\rangle}}\right)\label{equ34ver2}\\
& q^{\mu}= \lambda\left(Du^{\mu}+\beta\nabla^{\mu}T-4\omega^{\mu\nu}u_{\nu}-T\mathcal{B^{\langle\mu\rangle}}\right)\label{equ35ver2}\\
& \pi^{\mu\nu}=2\eta\left(\sigma^{\mu\nu}+T\mathcal{C^{\langle\mu\nu\rangle}}\right)\label{equ36ver2}\\
& \phi^{\mu\nu}=\gamma\left(\Omega^{\mu\nu}+2\beta\omega^{\langle\mu\rangle\langle\nu\rangle}+\mathcal{E^{\langle[\mu\nu]\rangle}}\right).\label{equ37ver2}
\end{align}
Analogous relations for various dissipative currents appearing in the spin tensor can be expressed as, 
\begin{align}
& \Phi=\chi_{1}\left(-2 u^{\alpha}\nabla^{\beta}(\beta\omega_{\alpha\beta})+\mathcal{F}\right)\label{equ38ver2}\\
& \tau^{\mu\beta}_{(s)}=\chi_{2}\left[-u^{\alpha}\left(\Delta^{\gamma\mu}\Delta^{\rho\beta}+\Delta^{\gamma\beta}\Delta^{\rho\mu}-\frac{2}{3}\Delta^{\gamma\rho}\Delta^{\mu\beta}\right)\nabla_{\gamma}(\beta\omega_{\alpha\rho})+\mathcal{G}^{\langle\mu\beta\rangle}\right]\label{equ39ver2}\\
& \tau^{\mu\beta}_{(a)}=\chi_{3} \left[-u^{\alpha}(\Delta^{\gamma\mu}\Delta^{\rho\beta}-\Delta^{\gamma\beta}\Delta^{\rho\mu})\nabla_{\gamma}(\beta\omega_{\alpha\rho})
   +\mathcal{H}^{\langle[\mu\beta]\rangle}\right]\label{equ40ver2}\\
 & \Theta^{\mu\alpha\beta}= -\chi_4  \left[-\Delta^{\delta\alpha}\Delta^{\rho\beta}\Delta^{\gamma\mu}\nabla_{\gamma}(\beta\omega_{\delta\rho})+\mathcal{I}^{\langle\mu\rangle\langle\alpha\rangle\langle\beta\rangle}\right]\label{equ41ver2}.
\end{align}
Here $\chi_1, \chi_2, \chi_{2},$ and $\chi_{4}$ are new spin-transport coefficients~\footnote{It is worth mentioning that expressions similar to the first terms appearing on the right-hand side in  Eqs.~\eqref{equ38ver2}-\eqref{equ41ver2} were also obtained using the first-order spin hydrodynamic approach considering the spin chemical potential leading order ($\mathcal{O}(1)$) in hydrodynamic gradient expansion~\cite{She:2021lhe}. However, such formalism~\cite{She:2021lhe}, is intrinsically different from our approach of formulating the first-order spin hydrodynamics~\cite{Daher:2022wzf,Daher:2022xon,Biswas:2022bht} as we consider $\omega_{\mu\nu}\sim \mathcal{O}(\partial)$. In our formalism, the derivative corrections to the spin tensor Eqs.~\eqref{equ38ver2}-\eqref{equ41ver2} don't contribute to the Navier-Stokes theory. This is one of the novel features of our spin hydrodynamic theory, where the nontrivial contribution of the dissipative parts of the spin tensor starts to contribute to the entropy current at the second order and beyond. Finally to avoid any confusion, Eqs.~\eqref{equ38ver2}-\eqref{equ41ver2} are dynamical equations, i.e, they contain the currents and their spacetime derivatives. Therefore, one should not get confused with the naive gradient counting at this level where all terms of Eqs.~\eqref{equ38ver2}-\eqref{equ41ver2} are not of second-order in gradients. As we will see later, this feature will allow us to recover the Naiver-Stokes spin-hydrodynamic equations~\cite{Daher:2022wzf,Daher:2022xon,Biswas:2022bht}.}. Using Eqs.~\eqref{equ33ver2}-\eqref{equ41ver2}, in Eq.~\eqref{equ32ver2} we obtain the following condition, 
\begin{align}
& -\frac{\beta}{\kappa}h^{\mu}h_{\mu}-\frac{\beta}{\lambda}q^{\mu}q_{\mu}+\frac{\beta}{2\eta}\pi^{\mu\nu}\pi_{\mu\nu}+\frac{\beta}{\zeta}\Pi^2+\frac{1}{\gamma}\phi^{\mu\nu}\phi_{\mu\nu}\nonumber\\
& +\frac{1}{\chi_1}\Phi^2+\frac{1}{\chi_2}\tau^{\mu\nu}_{(s)}\tau_{\mu\nu(s)}+\frac{1}{\chi_3}\tau^{\mu\nu}_{(a)}\tau_{\mu\nu(a)}-\frac{1}{\chi_4}\Theta^{\mu\alpha\beta}\Theta_{\mu\alpha\beta}\geq 0.
\end{align}
This immediately implies that $\kappa\geq 0$, $\lambda\geq 0$, $\eta\geq0$, $\zeta\geq0$ $\gamma\geq0$, $\chi_1\geq 0$,
$\chi_2\geq 0$, $\chi_3\geq 0$, and $\chi_4\geq 0$. We emphasize that the presence of $\mathcal{D}$, $\mathcal{A^{\langle\mu\rangle}}$, $\mathcal{B^{\langle\mu\rangle}}$, $\mathcal{C^{\langle\mu\nu\rangle}}$ and $\mathcal{E^{\langle[\mu\nu]\rangle}}$ in Eqs.~\eqref{equ33ver2}-\eqref{equ37ver2} shows that for the second order theory the constitutive relations of $\Pi$, $h^{\mu}$, $q^{\mu}$, $\pi^{\mu\nu}$ and $\phi^{\mu\nu}$ are not simply expressed by Eqs.~\eqref{eq11}-\eqref{eq15} in terms of basic hydrodynamic variables, $T$, $u^{\mu}$ and $\omega^{\mu\nu}$. Therefore in the second order theory $\Pi$, $h^{\mu}$, $q^{\mu}$, $\pi^{\mu\nu}$ and $\phi^{\mu\nu}$ should be considered as independent hydrodynamic variables along with $T$, $u^{\mu}$ and $\omega^{\mu\nu}$. The evolution equation of new hydrodynamic variables can be obtained from Eqs.~\eqref{equ33ver2}-\eqref{equ37ver2}. Using explicit expressions of $\mathcal{D}$, $\mathcal{A^{\langle\mu\rangle}}$, $\mathcal{B^{\langle\mu\rangle}}$, $\mathcal{C^{\langle\mu\nu\rangle}}$ and $\mathcal{E^{\langle[\mu\nu]\rangle}}$ we can write the evolution equations of different dissipative currents as,
\begin{align}
D\Pi+\frac{\Pi}{\tau_{\Pi}}=&-\frac{1}{2a_1}\bigg[\beta\theta+a_1\Pi\theta+\Pi Da_1+(1-l_{\Pi h})h^{\mu}\nabla_{\mu}b_1-b_1(1-\tilde{l}_{\Pi h})h^{\mu}Du_{\mu}+b_1\nabla_{\mu}h^{\mu}+l_{\Pi q}q^{\mu}\nabla_{\mu}b_4
\nonumber\\
& -\tilde{l}_{\Pi q}b_4  q^{\mu}Du_{\mu}+b_4 \nabla_{\mu}q^{\mu}+l_{\Theta\Pi}\Theta^{\alpha\mu\nu}\Delta_{\alpha\mu}\nabla_{\nu}c_3-\tilde{l}_{\Theta\Pi}c_3\Delta_{\alpha\mu}\Theta^{\alpha\mu\nu}Du_{\nu}+c_3\Delta_{\alpha\beta}\nabla_{\mu}\Theta^{\alpha\beta\mu}\bigg],
\label{equ43ver2}
\end{align}
\begin{align}
Dh^{\langle\mu\rangle}+\frac{h^{\mu}}{\tau_{h}} =&-\frac{1}{2a_3}\bigg[\beta(Du^{\mu}-\beta\nabla^{\mu}T)+a_3h^{\mu}\theta+h^{\mu}Da_3+l_{\Pi h}\Pi\nabla^{\mu}b_1+b_1\nabla^{\mu}\Pi-b_1\tilde{l}_{\Pi h}\Pi Du^{\mu}+l_{\pi h}\pi^{\lambda\mu}\nabla_{\lambda}b_2
\nonumber\\
&+b_2\Delta^{\mu}_{~\nu}\nabla_{\lambda}\pi^{\lambda\nu}-b_2\tilde{l}_{\pi h}\pi^{\lambda\mu}Du_{\lambda}+l_{\phi h}\phi^{\lambda\mu}\nabla_{\lambda}b_3+b_3\Delta^{\mu}_{~\nu}\nabla_{\lambda}\phi^{\lambda\nu}-b_3\tilde{l}_{\phi h}\phi^{\lambda\mu}Du_{\lambda}+l_{\Phi h}\Phi\nabla^{\mu}\tilde{b}_1\nonumber\\
& +\tilde{b}_1\nabla^{\mu}\Phi-\tilde{b}_1\tilde{l}_{\Phi h}\Phi Du^{\mu}+l_{\tau_s h}\tau^{\lambda\mu}_{(s)}\nabla_{\lambda}\tilde{b}_2+\tilde{b}_2\Delta^{\mu}_{~\nu}\nabla_{\lambda}\tau^{\lambda\nu}_{(s)}-\tilde{b}_2\tilde{l}_{\tau_{s}h}\tau^{\lambda\mu}_{(s)}Du_{\lambda}+l_{\tau_a h}\tau^{\lambda\mu}_{(a)}\nabla_{\lambda}\tilde{b}_3\nonumber\\
&+\tilde{b}_3\Delta^{\mu}_{~\nu}\nabla_{\lambda}\tau^{\lambda\nu}_{(a)}-\tilde{b}_3\tilde{l}_{\tau_{a}h}\tau^{\lambda\mu}_{(a)}Du_{\lambda}\bigg],
\label{equ44ver2}
\end{align}
\begin{align}
Dq^{\langle\mu\rangle}+\frac{q^{\mu}}{\tau_{q}} =&\frac{1}{2a_4}\bigg[\beta(\beta\nabla^{\mu}T+Du^{\mu}-4\omega^{\mu\nu}u_{\nu})-a_4 q^{\mu}\theta-q^{\mu}Da_4-(1-l_{\Pi q})\Pi\nabla^{\mu}b_4-b_4 \nabla^{\mu}\Pi
\nonumber\\
& +b_4(1-\tilde{l}_{\Pi q}) \Pi Du^{\mu}-(1-l_{\pi q}) \pi^{\lambda\mu}\nabla_{\lambda}b_5-b_5\Delta^{\mu}_{~\nu}\nabla_{\lambda}\pi^{\lambda\nu}+b_5(1-\tilde{l}_{\pi q}) \pi^{\lambda\mu}Du_{\lambda}-l_{\phi q} \phi^{\lambda\mu}\nabla_{\lambda}b_6
\nonumber\\
& -b_6 \Delta^{\mu}_{~\nu}\nabla_{\lambda}\phi^{\lambda\nu}+b_6\tilde{l}_{\phi q} \phi^{\lambda\mu}Du_{\lambda}-l_{\Phi q} \Phi\nabla^{\mu}\tilde{b}_4-\tilde{b}_4 \nabla^{\mu}\Phi+\tilde{b}_4\tilde{l}_{\Phi q} \Phi Du^{\mu}-l_{\tau_s q} \tau^{\lambda\mu}_{(s)}\nabla_{\lambda}\tilde{b}_5
\nonumber\\
& -\tilde{b}_5 \Delta^{\mu}_{~\nu}\nabla_{\lambda}\tau^{\lambda\nu}_{(s)}+\tilde{b}_5\tilde{l}_{\tau_{s}q} \tau^{\lambda\mu}_{(s)}Du_{\lambda}-l_{\tau_a q} \tau^{\lambda\mu}_{(a)}\nabla_{\lambda}\tilde{b}_6-\tilde{b}_6 \Delta^{\mu}_{~\nu}\nabla_{\lambda}\tau^{\lambda\nu}_{(a)}+\tilde{b}_6\tilde{l}_{\tau_{a}q} \tau^{\lambda\mu}_{(a)}Du_{\lambda}\bigg],
\label{equ45ver2}
\end{align}
\begin{align}
D\pi^{\langle\mu\nu\rangle}+\frac{\pi^{\mu\nu}}{\tau_{\pi}}=&-\frac{1}{2a_2}\bigg[\beta\sigma^{\mu\nu}+a_2 \theta \pi^{\mu\nu}+ \pi^{\mu\nu}Da_2+(1-l_{\pi h}) h^{\langle\mu}\nabla^{\nu\rangle}b_2-b_2(1-\tilde{l}_{\pi h})h^{\langle\mu}Du^{\nu\rangle}
\nonumber\\
&
+b_2\nabla^{\langle\mu}h^{\nu\rangle}+l_{\pi q}q^{\langle\mu}\nabla^{\nu\rangle}b_5-\tilde{l}_{\pi q}b_5 q^{\langle\mu}Du^{\nu\rangle}+b_5 \nabla^{\langle\mu}q^{\nu\rangle}+l_{\Theta\pi}\Theta^{\langle\mu\nu\rangle\alpha}\nabla_{\alpha}c_4
\nonumber\\
&-\tilde{l}_{\Theta\pi}c_4\Theta^{\langle\mu\nu\rangle\alpha}Du_{\alpha}+c_4\nabla_{\alpha}\Theta^{\langle\mu\nu\rangle\alpha}\bigg],
\label{equ46ver2}
\end{align}
\begin{align}
D\phi^{\langle[\mu\nu]\rangle}+\frac{\phi^{\mu\nu}}{\tau_{\phi}} =&-\frac{1}{2a_5}\bigg[\left(\Omega^{\mu\nu}+2\beta\omega^{\langle\mu\rangle\langle\nu\rangle}\right)+a_5\theta\phi^{\mu\nu}+\phi^{\mu\nu}Da_5+(1-l_{\phi h})h^{[\nu}\nabla^{\mu]}b_3
\nonumber\\
&-b_3(1-\tilde{l}_{\phi h})h^{[\nu}Du^{\mu]}
+b_3\Delta^{[\mu\nu]}_{[\alpha\beta]}\nabla^{[\alpha}h^{\beta]}+(1-l_{\phi q})q^{[\nu}\nabla^{\mu]}b_6-b_6(1-\tilde{l}_{\phi q})q^{[\nu}Du^{\mu]}
\nonumber\\
&
+b_6\Delta^{[\mu\nu]}_{[\alpha\beta]}\nabla^{[\alpha}q^{\beta]}+l_{\Theta\phi}\Theta^{\lambda\mu\nu}\nabla_{\lambda}c_1-\tilde{l}_{\Theta\phi}c_1\Theta^{\lambda\mu\nu}Du_{\lambda}+c_3\Delta^{[\mu\nu]}_{[\alpha\beta]}\nabla_{\lambda}\Theta^{\lambda\alpha\beta}+k_{\Theta\phi}\Theta^{[\mu\nu]\lambda}\nabla_{\lambda}c_7
\nonumber\\
&-\tilde{k}_{\Theta\phi}c_7\Theta^{[\mu\nu]\lambda}Du_{\lambda}+c_7\Delta^{[\mu\nu]}_{[\alpha\beta]}\nabla_{\lambda}\Theta^{[\alpha\beta]\lambda}\bigg].
 \label{equ47ver2}
\end{align}
In the above equations, the relaxation times of various dissipative quantities are defined as, $\tau_{\Pi}=-2a_1\zeta T\geq 0$, $\tau_{h}=2a_3\kappa T\geq 0$, $\tau_{q}=2a_4\lambda T\geq 0$, $\tau_{\pi}=-4a_2\eta T\geq 0$, and $\tau_{\phi}=-2a_5\gamma\geq 0$. Moreover we define, $Dh^{\langle\mu\rangle}=\Delta^{\mu}_{~\nu}Dh^{\nu}$, $Dq^{\langle\mu\rangle}=\Delta^{\mu}_{~\nu}Dq^{\nu}$, $D\pi^{\langle\mu\nu\rangle}=\Delta^{\mu\nu}_{\alpha\beta}D\pi^{\alpha\beta}$, and $D\phi^{\langle[\mu\nu]\rangle}=\Delta^{[\mu\nu]}_{[\alpha\beta]}D\phi^{\alpha\beta}$. The dissipative currents appearing in the spin tensor also satisfy similar relaxation type equations,
\begin{align}
D\Phi+\frac{\Phi}{\tau_{\Phi}}=&-\frac{1}{2\tilde{a}_1}\bigg[-2 u^{\alpha}\nabla^{\beta}(\beta\omega_{\alpha\beta})+\tilde{a}_1\theta\Phi+\Phi D\tilde{a}_1+(1-l_{\Phi h})h^{\mu}\nabla_{\mu}\tilde{b}_1-(1-\tilde{l}_{\Phi h})\tilde{b}_1h^{\mu}Du_{\mu}+\tilde{b}_1\nabla_{\mu}h^{\mu}\nonumber\\
&+(1-l_{\Phi h})q^{\mu}\nabla_{\mu}\tilde{b}_4-(1-\tilde{l}_{\Phi q})\tilde{b}_4 q^{\mu}Du_{\mu}+\tilde{b}_4\nabla_{\mu}q^{\mu}+l_{\Theta\Phi}\Theta^{\alpha\mu\nu}\Delta_{\alpha\mu}\nabla_{\nu}c_5
\nonumber\\
&-\tilde{l}_{\Theta\Pi}c_5\Delta_{\alpha\mu}\Theta^{\alpha\mu\nu}Du_{\nu}+c_5\Delta_{\alpha\beta}\nabla_{\mu}\Theta^{\alpha\beta\mu}\bigg],
\label{equ48ver2}
\end{align}
\begin{align}
D\tau^{\langle\mu\nu\rangle}_{(s)}+\frac{\tau^{\mu\nu}_{(s)}}{\tau_{\tau_s}}  =&-\frac{1}{2\tilde{a}_2}\bigg[-u^{\alpha}\left(\Delta^{\gamma\mu}\Delta^{\rho\nu}+\Delta^{\gamma\nu}\Delta^{\rho\mu}-\frac{2}{3}\Delta^{\gamma\rho}\Delta^{\mu\nu}\right)\nabla_{\gamma}(\beta\omega_{\alpha\rho})+\tilde{a}_2\theta \tau^{\mu\nu}_{(s)}+\tau^{\mu\nu}_{(s)}D\tilde{a}_2\nonumber\\
&+(1-l_{\tau_{s} h})h^{\langle\mu}\nabla^{\nu\rangle}\tilde{b}_2-\tilde{b}_2(1-\tilde{l}_{\tau_s h})h^{\langle\mu}Du^{\nu\rangle}+\tilde{b}_2\nabla^{\langle\mu}h^{\nu\rangle}+(1-l_{\tau_s q})q^{\langle\mu}\nabla^{\nu\rangle}\tilde{b}_5
\nonumber\\
&-(1-\tilde{l}_{\tau_s q})\tilde{b}_5 q^{\langle\mu}Du^{\nu\rangle}+\tilde{b}_5\nabla^{\langle\mu}q^{\nu\rangle}+l_{\Theta\tau_s}\Theta^{\langle\mu\nu\rangle\lambda}\nabla_{\lambda}c_6-\tilde{l}_{\Theta\tau_s}c_6\Theta^{\langle\mu\nu\rangle\lambda}Du_{\lambda}+c_6\nabla_{\lambda}\Theta^{\langle\mu\nu\rangle\lambda}\bigg],
\label{equ49ver2}
\end{align}
\begin{align}
D\tau^{\langle[\mu\nu]\rangle}_{(a)}+\frac{\tau^{\mu\nu}_{(a)}}{\tau_{\tau_a}}=&-\frac{1}{2\tilde{a}_3}\bigg[-u^{\alpha}(\Delta^{\gamma\mu}\Delta^{\rho\nu}-\Delta^{\gamma\nu}\Delta^{\rho\mu})\nabla_{\gamma}(\beta\omega_{\alpha\rho})   +\tilde{a}_3\theta\tau^{\mu\nu}_{(a)}+\tau^{\mu\nu}_{(a)}D\tilde{a}_3+(1-l_{\tau_a h})h^{[\nu}\nabla^{\mu]}\tilde{b}_3\nonumber\\
&-\tilde{b}_3(1-\tilde{l}_{\tau_a h})h^{[\nu}Du^{\mu]}+\tilde{b}_3\Delta^{[\mu\nu]}_{[\alpha\beta]}\nabla^{[\alpha}h^{\beta]}+(1-l_{\tau_a q})q^{[\nu}\nabla^{\mu]}\tilde{b}_6-\tilde{b}_6(1-\tilde{l}_{\tau_a q})q^{[\nu}Du^{\mu]}\nonumber\\
&+\tilde{b}_6\Delta^{[\mu\nu]}_{[\alpha\beta]}\nabla^{[\alpha}q^{\beta]}+l_{\Theta\tau_a}\Theta^{\lambda\mu\nu}\nabla_{\lambda}c_2-\tilde{l}_{\Theta\tau_a}c_2\Theta^{\lambda\mu\nu}Du_{\lambda}+c_2\Delta^{[\mu\nu]}_{[\alpha\beta]}\nabla_{\lambda}\Theta^{\lambda\alpha\beta}+k_{\Theta\tau_a}\Theta^{[\mu\nu]\lambda}\nabla_{\lambda}c_8\nonumber\\
&-\tilde{k}_{\Theta\tau_a}c_8\Theta^{[\mu\nu]\lambda}Du_{\lambda}+c_8\Delta^{[\mu\nu]}_{[\alpha\beta]}\nabla_{\lambda}\Theta^{[\alpha\beta]\lambda}\bigg],
\label{equ50ver2}
\end{align}
\begin{align}
D\Theta^{\langle\alpha\rangle\langle\mu\rangle\langle\nu\rangle}+\frac{\Theta^{\alpha\mu\nu}}{\tau_{\Theta}}=&-\frac{1}{2\tilde{a}_4}\bigg[-\Delta^{\delta\mu}\Delta^{\rho\nu}\Delta^{\gamma\alpha}\nabla_{\gamma}(\beta\omega_{\delta\rho})+\tilde{a}_4\theta\Theta^{\alpha\mu\nu}+ \Theta^{\alpha\mu\nu}D\tilde{a}_4+ (1-l_{\Theta \phi})\phi^{\mu\nu}\nabla^{\alpha}c_1\nonumber\\
& -(1-\tilde{l}_{\Theta \phi})c_1\phi^{\mu\nu}Du^{\alpha}+c_1\Delta^{\alpha a}\Delta^{\mu b}\Delta^{\nu  c}\nabla_{a}\phi_{bc}+ (1-l_{\Theta \tau_a})\tau_{(a)}^{\mu\nu}\nabla^{\alpha}c_2\nonumber\\
& -(1-\tilde{l}_{\Theta \tau_a})c_2\tau^{\mu\nu}_{(a)}Du^{\alpha}+c_2\Delta^{\alpha a}\Delta^{\mu b}\Delta^{\nu  c}\nabla_a\tau_{bc(a)}+(1-l_{\Theta\Pi})\Pi\Delta^{\alpha[\mu}\nabla^{\nu]}c_3\nonumber\\
& -(1-\tilde{l}_{\Theta\Pi})c_3\Pi\Delta^{\alpha[\mu}Du^{\nu]}+c_3\Delta^{\alpha[\mu}\nabla^{\nu]}\Pi+(1-l_{\Theta\Phi})\Phi\Delta^{\alpha[\mu}\nabla^{\nu]}c_5\nonumber\\
& -(1-\tilde{l}_{\Theta\Phi})c_5\Phi\Delta^{\alpha[\mu}Du^{\nu]}+c_5\Delta^{\alpha[\mu}\nabla^{\nu]}\Phi+(1-l_{\Theta\pi})\pi^{\alpha[\mu}\nabla^{\nu]}c_4
\nonumber\\
& -(1-\tilde{l}_{\Theta\pi})c_4\pi^{\alpha[\mu}Du^{\nu]}+c_4\Delta^{\alpha a}\Delta^{\mu b} \Delta^{\nu c}\nabla_{[c}\pi_{ab]}+(1-l_{\Theta\tau_s})\tau_{(s)}^{\alpha[\mu}\nabla^{\nu]}c_6\nonumber\\
& -(1-\tilde{l}_{\Theta\tau_s})c_6\tau_{(s)}^{\alpha[\mu}Du^{\nu]}+c_6\Delta^{\alpha a}\Delta^{\mu b} \Delta^{\nu c}\nabla_{[c}\tau_{(s)ab]}+(1-k_{\Theta\phi})\phi^{\alpha[\mu}\nabla^{\nu]}c_7
\nonumber\\
& -(1-\tilde{k}_{\Theta\phi})c_7\phi^{\alpha[\mu}Du^{\nu]}+c_7\Delta^{\alpha a}\Delta^{\mu b} \Delta^{\nu c}\nabla_{[c}\phi_{ab]}+(1-k_{\Theta\tau_a})\tau_{(a)}^{\alpha[\mu}\nabla^{\nu]}c_8\nonumber\\
& -(1-\tilde{k}_{\Theta\tau_a})c_8\tau_{(a)}^{\alpha[\mu}Du^{\nu]}+c_8\Delta^{\alpha a}\Delta^{\mu b} \Delta^{\nu c}\nabla_{[c}\tau_{(a)ab]}\bigg].
\label{equ51ver2}
\end{align}
In above equations, $D\tau^{\langle\mu\nu\rangle}_{(s)}\equiv \Delta^{\mu\nu}_{\alpha\beta}D\tau_{(s)}^{\alpha\beta}$, $D\tau^{\langle[\mu\nu]\rangle}_{(a)}\equiv \Delta^{[\mu\nu]}_{[\alpha\beta]}D\tau_{(a)}^{\alpha\beta}$,  $D\Theta^{\langle\alpha\rangle\langle\mu\rangle\langle\nu\rangle}\equiv \Delta^{\alpha a}\Delta^{\mu b}\Delta^{\nu  c}D\Theta_{abc} $. Various spin-relaxation times can be identified as, $\tau_{\Phi}=-2\tilde{a_1}\chi_1\geq 0$, $\tau_{\tau_s}\equiv -2\tilde{a}_2\chi_2\geq 0$, $\tau_{\tau_a}\equiv -2\tilde{a}_3\chi_3\geq 0$, and $\tau_{\Theta}\equiv 2\tilde{a}_4\chi_4\geq 0$. 
In comparison to the first-order spin hydrodynamics, one of the most important features of the second-order theory is the presence of relaxation times corresponding to various the dissipative currents. The time scales within which dissipative currents respond to hydrodynamic gradients are represented by these relaxation times. These relaxation times are expected to make the second-order theory free from any problems appearing from acausality and hydrodynamic instability. However, such an important feature of the second-order theory comes at a price. With respect to the first-order theory, there are more parameters or transport coefficients in the second-order theory. Although Eqs.~\eqref{equ43ver2}-\eqref{equ47ver2} and Eqs.~\eqref{equ48ver2}-\eqref{equ51ver2} are all relaxation type equations, but there is a striking difference between these equations. Note that Eqs.~\eqref{equ43ver2}-\eqref{equ47ver2} contain terms of $\mathcal{O}(\partial)$ and $\mathcal{O}(\partial^2)$ on both sides. But this is not true for Eqs.~\eqref{equ48ver2}-\eqref{equ51ver2}. The right hand sides of Eqs.~\eqref{equ48ver2}-\eqref{equ51ver2} does not contain terms of the order $\mathcal{O}(\partial)$. This is because the dissipative parts of the spin tensor do not contribute to the entropy current analysis at the Navier-Stokes limit where all the dissipative currents are expressed as $\mathcal{O}(\partial)$ terms. To check the consistency of the formalism, it is natural to look for the Navier-Stokes limit of the second-order theory. This can be achieved by ignoring all second-order terms in the hydrodynamic gradient expansion in Eqs.~\eqref{equ43ver2}-\eqref{equ51ver2}. In this limit, we retrieve back the constitutive relation of various dissipative currents associated with the energy-momentum tensor, e.g., from Eq.~\eqref{equ43ver2} we find, after ignoring all $\mathcal{O}(\partial^2)$ terms, 
\begin{align}
\Pi=-\frac{\tau_{\Pi}}{2a_1}\beta\theta=\zeta\theta.
\end{align}
Similarly constitutive relations for $h^{\mu}$, $q^{\mu}$, $\pi^{\mu\nu}$, $\phi^{\mu\nu}$ can be obtained from Eqs.~\eqref{equ44ver2}, \eqref{equ45ver2}, \eqref{equ46ver2}, and \eqref{equ47ver2} respectively. These expressions will match Eqs.~\eqref{eq11}-\eqref{eq15}. However if we ignore all $\mathcal{O}(\partial^2)$ terms in Eqs.~\eqref{equ48ver2}-\eqref{equ51ver2} then we observe that $\Phi= 0+\mathcal{O}(\partial^2)$, $\tau^{\mu\nu}_{(s)}= 0+\mathcal{O}(\partial^2)$, $\tau^{\mu\nu}_{(a)}= 0+\mathcal{O}(\partial^2)$, $\Theta^{\alpha\mu\nu}= 0+\mathcal{O}(\partial^2)$. This immediately implies that at the Navier-Stokes limit gradient correction terms to the spin tensor do not contribute to the entropy production, and $S^{\alpha\mu\nu}_{(1)}$ can only be obtained for the second order theory.
%%%%%%%%%%%%%%%%%%%%%%%%%%%%%%%%%%%%%%%%%%%%

%%%%%%%%%%%%%%%%%%%%%%%%%%%%%%%%%%%%%%%%%%%%
\section{Conclusions and outlook}
\label{secV}
In this paper, we show a new derivation of the second-order dissipative spin hydrodynamic equations. This formulation is based on the positivity of the entropy production for a dissipative system. We consider an energy-momentum tensor which is asymmetric and the spin tensor has a simple phenomenological form where it is only anti-symmetric in the last two indices. 
%One of the important features of this framework is that it can be applied to a system composed of massive as well as massless spin half particles.
One can retrieve the correct Navier-Stokes limit as well as global equilibrium conditions.  Our calculations can be used to study macroscopic spin evolution and possibly it will help us to solve the puzzle related to the longitudinal polarization of Lambda particles in a dynamical way. But this requires a proper numerical implementation of spin hydrodynamic equations along with appropriate initial conditions and hadronic freezeout. One immediate future task would be to study the stability and causality analysis to pin down the region of applicability of this theory. Although we have obtained relaxation time like hydrodynamic equations it lacks a proper understanding of the microscopic theory. This is manifested in large numbers of unknown transport coefficients and relaxation times. Note that a dissipative hydrodynamic theory captures the long wavelength and long-time behavior of a system away from equilibrium. On the other hand transport coefficients encodes the microscopic physics at a length scale smaller than the domain of applicability of hydrodynamics. The estimation of various relaxation times and transport coefficients is very important for phenomenological applications. Only a bottom-up approach to spin-hydrodynamic where one obtains a spin-hydrodynamic equation using a kinetic theory approach can bridge this problem. Finding an equivalent kinetic theory approach without further assumptions will be a good direction to explore as a future task. 

 % \medskip
%%%%%%%%%%%%%%%%%%%%%%%%%%%%%%%%%%%%%%%%%%%%
{\bf Acknowledgements:} We thank Leonardo Tinti for clarifications. This work was supported in part by the Polish National Science Centre Grant Nos 2018/30/E/ST2/00432 and 2020/39/D/ST2/02054. RB acknowledges the financial support from SPS, NISER (Bhubaneswar, India)  planned project RIN4001. RB has been supported, in part, by the Polish National Science Centre (NCN) Sonata Bis grant 2019/34/E/ST3/00405 and the International Max Planck Research School for ``Quantum Dynamics and Control''.
%%%%%%%%%%%%%%%%%%%%%%%%%%%%%%%%%%%%%%%%%%%%
\appendix
\section{Constraint on the form of $Q^{\mu}$}
\label{app1}
Contracting the second-order entropy current~\eqref{eq16} with the fluid four-velocity, and using the fact that $u_{\mu}S^{\mu\alpha\beta}_{(1)}=0$, we get  
\begin{align}
u_{\mu}s^{\mu}_{\rm IS}&=u_{\mu}s^{\mu}_{\rm NS}+u_{\mu}Q^{\mu}.
\end{align}
Substituting the form of $s_{\rm NS}^{\mu}$~\eqref{eq10} in the above equation, we have
\begin{align}
u_{\mu}s^{\mu}_{\rm IS}&=u_{\mu}\left[s^{\mu}+\beta_{\nu}T^{\mu\nu}_{(1)}+\mathcal{O}(\partial^{2})\right]+u_{\mu}Q^{\mu},\nonumber\\
&=u_{\mu}s^{\mu}+u_{\mu}Q^{\mu}.
\end{align}
Utilizing the perfect-fluid energy-momentum tensor~\eqref{perfect fluid EMT}, and replacing the form of the entropy current $s^{\mu}$~\eqref{eq:9} we find,
\begin{align}
u_{\mu}s^{\mu}_{\rm IS}&=u_{\mu}\left(\beta_{\nu}T^{\mu\nu}_{(0)}+\beta^{\mu}p-\beta^{\mu}\omega_{\alpha\beta}S^{\alpha\beta}\right)+u_{\mu}Q^{\mu},\nonumber\\
&=u_{\mu}\left[\beta_{\nu}(\varepsilon+p)u^{\mu}u^{\nu}-\beta_{\nu}pg^{\mu\nu}+\beta^{\mu}p-\beta^{\mu}\omega_{\alpha\beta}S^{\alpha\beta}\right]+u_{\mu}Q^{\mu},\nonumber\\
&=\beta\left[(\varepsilon+p)-\omega_{\alpha\beta}S^{\alpha\beta}\right]+u_{\mu}Q^{\mu}.
\end{align}
Finally, using the generalized first law of thermodynamics~\eqref{therodynamiceq}, we obtain
\begin{align}
u_{\mu}s^{\mu}_{\rm IS}=s+u_{\mu}Q^{\mu}.
\end{align}
Employing the fact that entropy is maximum in equilibrium, we obtain the constraint on $Q^{\mu}$, i.e,
\begin{align}
u_{\mu}Q^{\mu}\leq 0.
\end{align}

%%%%%%%%%%%%%%%%%%%%%%%%%%%%%%%%%%%%%%%%%%%%%%%%%%%%%%%

\section{Decomposition of an arbitrary 3-rank tensor antisymmetric in last two indices}
\label{appendixH}
Let us consider an arbitrary three-rank tensor $\phi^{\lambda\mu\nu}$ antisymmetric in last two indices. Employing the decomposition of its first index into the parts transverse and parallel to four-velocity, one has
\begin{align}
\phi^{\lambda\mu\nu}& =g^{\lambda}_{~\alpha} \phi^{\alpha\mu\nu}=(u^{\lambda}u_{\alpha}+\Delta^{\lambda}_{~\alpha})\phi^{\alpha\mu\nu}\nonumber\\
& = u^{\lambda}\gamma^{\mu\nu}+\Delta^{\lambda}_{~\alpha}\phi^{\alpha\mu\nu}\nonumber\\
& = u^{\lambda}\gamma^{\mu\nu}+\phi^{\langle\lambda\rangle\mu\nu}
\label{equ88new1}
\end{align}
Here we define antisymmetric tensor $\gamma^{\mu\nu}\equiv u_{\alpha}\phi^{\alpha\mu\nu}$. This immediately implies that $F^{\nu}\equiv u_{\mu}\gamma^{\mu\nu}$ satisfies $F\cdot u=0$. In the next step, we proceed with the decomposition of $\gamma^{\mu\nu}$
\begin{align}
\gamma^{\mu\nu}& =g^{\mu}_{~\rho}\gamma^{\rho\nu}=(u^{\mu}u_{\rho}+\Delta^{\mu}_{~\rho})\gamma^{\rho\nu}=u^{\mu}F^{\nu}+\gamma^{\langle\mu\rangle\nu}\nonumber\\
& = u^{\mu}F^{\nu}+g^{\nu}_{~\rho}\gamma^{\langle\mu\rangle\rho}=u^{\mu}F^{\nu}+(u^{\nu}u_{\rho}+\Delta^{\nu}_{~\rho})\gamma^{\langle\mu\rangle\rho}\nonumber\\
& = u^{\mu}F^{\nu}+u^{\nu}u_{\rho}\gamma^{\langle\mu\rangle\rho}+\gamma^{\langle\mu\rangle\langle\nu\rangle}
\label{equ89new1}
\end{align}
It can be easily shown that $u^{\nu}u_{\rho}\gamma^{\langle\mu\rangle\rho}=-u^{\nu}F^{\mu}$. Therefore, $\gamma^{\mu\nu}$ has the form, 
\begin{align}
\gamma^{\mu\nu}=u^{\mu}F^{\nu}-u^{\nu}F^{\mu}+\gamma^{\langle\mu\rangle\langle\nu\rangle}.
\end{align}
Now, let us consider the last term in Eq.~\eqref{equ88new1}, 
\begin{align}
\phi^{\langle\lambda\rangle\mu\nu}=g^{\mu}_{~\rho}\phi^{\langle\lambda\rangle\rho\nu}=(u^{\mu}u_{\rho}+\Delta^{\mu}_{~\rho})\phi^{\langle\lambda\rangle\rho\nu}=u^{\mu}u_{\rho}\phi^{\langle\lambda\rangle\rho\nu}+\phi^{\langle\lambda\rangle\langle\mu\rangle\nu}
\end{align}
Defining $u_{\rho}\phi^{\langle\lambda\rangle\rho\nu}\equiv-\Sigma^{\lambda\nu}$, implies $u_{\lambda}\Sigma^{\lambda\nu}=0$. Therefore, 
\begin{align}
\phi^{\langle\lambda\rangle\mu\nu} & = \phi^{\langle\lambda\rangle\langle\mu\rangle\nu}-u^{\mu}\Sigma^{\lambda\nu}\nonumber\\
& = g^{\nu}_{~\alpha}\phi^{\langle\lambda\rangle\langle\mu\rangle\alpha}-u^{\mu}\Sigma^{\lambda\nu}\nonumber\\
& = \phi^{\langle\lambda\rangle\langle\mu\rangle\langle\nu\rangle}+u^{\nu}u_{\alpha}\phi^{\langle\lambda\rangle\langle\mu\rangle\alpha}-u^{\mu}\Sigma^{\lambda\nu}\nonumber\\
& = \phi^{\langle\lambda\rangle\langle\mu\rangle\langle\nu\rangle}+u^{\nu}\Sigma^{\lambda\mu}-u^{\mu}\Sigma^{\lambda\nu}.
\label{equ92new1}
\end{align}
Using Eqs.~\eqref{equ89new1} and \eqref{equ92new1} in Eq.~\eqref{equ88new1} we obtain, 
\begin{align}
\phi^{\lambda\mu\nu}=u^{\lambda}\left(u^{\mu}F^{\nu}-u^{\nu}F^{\mu}+\gamma^{\langle\mu\rangle\langle\nu\rangle}\right)+u^{\nu}\Sigma^{\lambda\mu}-u^{\mu}\Sigma^{\lambda\nu}+\phi^{\langle\lambda\rangle\langle\mu\rangle\langle\nu\rangle}.
\end{align}
Here we can introduce $\mathcal{S}^{\mu\nu}\equiv u^{\mu}F^{\nu}-u^{\nu}F^{\mu}+\gamma^{\langle\mu\rangle\langle\nu\rangle}$. Noticing that $\mathcal{S}^{\mu\nu}$ is an antisymmetric tensor that can also be decomposed as $\mathcal{S}^{\mu\nu}\equiv u^{\mu}\kappa^{\nu}-u^{\nu}\kappa^{\mu}+\epsilon^{\mu\nu\alpha\beta}u_{\alpha}\omega_{\beta}$, with $u\cdot\kappa=0$ and $u\cdot\omega=0$, we identify $F^{\nu}=\kappa^{\nu}$, and $\gamma^{\langle\mu\rangle\langle\nu\rangle}\equiv\epsilon^{\mu\nu\alpha\beta}u_{\alpha}\omega_{\beta}$~\cite{Huang:2011dc}. Since $\Sigma^{\mu\nu}$ is asymmetric (not antisymmetric!) and orthogonal to $u^{\mu}$ it can also be decomposed into symmetric ($\Sigma_{(s)}^{\mu\nu}$) and antisymmetric ($\Sigma_{(a)}^{\mu\nu}$) parts. The symmetric part can be further decomposed into a trace ($\Sigma$) and a traceless part ($\Sigma_s^{\langle\mu\nu\rangle}$). Finally, we obtain the following expression, 
\begin{align}
\phi^{\lambda\mu\nu}= u^{\lambda}\mathcal{S}^{\mu\nu}+\left(u^{\nu}\Delta^{\lambda\mu}-u^{\mu}\Delta^{\lambda\nu}\right)\Sigma+\left(u^{\nu}\Sigma_{(s)}^{\langle\lambda\mu\rangle}-u^{\mu}\Sigma_{(s)}^{\langle\lambda\nu\rangle}\right)+\left(u^{\nu}\Sigma_{(a)}^{\lambda\mu}-u^{\mu}\Sigma_{(a)}^{\lambda\nu}\right)+\phi^{\langle\lambda\rangle\langle\mu\rangle\langle\nu\rangle}.
\end{align}
One may check that the number of degrees of freedom (DOF) matches for the quantities on both sides of the above equation. The tensor $\phi^{\lambda\mu\nu}$ has in total 24 DOF. At the same time, $\mathcal{S}^{\mu\nu}$ has 6 DOF, and $\Sigma$ is a scalar, hence it has only one DOF. $\Sigma_{(s)}^{\langle\mu\nu\rangle}$ is symmetric, traceless, and orthogonal to the fluid flow vector, hence it has 5 DOF, while $\Sigma_{(a)}^{\mu\nu}$ is antisymmetric and transverse to the fluid flow, hence it has 3 DOF. Finally, $\phi^{\langle\lambda\rangle\langle\mu\rangle\langle\nu\rangle}$ is antisymmetric in the last two indices and orthogonal to flow vector in all indices, hence it has only 9 DOF.

\section{Derivation of Eq.~\eqref{eq24}}
\label{appendixB}
We start with the entropy current given in Eq.~\eqref{eq16},
\begin{align}
& s^{\mu}_{\rm IS}=\beta_{\nu}T^{\mu\nu}+p\beta^{\mu}-\beta\omega_{\alpha\beta}S^{\mu\alpha\beta}+Q^{\mu}\nonumber\\
\implies & \partial_{\mu}s^{\mu}_{\rm IS}= T^{\mu\nu}\partial_{\mu}\beta_{\nu}+\beta_{\nu}\partial_{\mu}T^{\mu\nu}+\partial_{\mu}(p\beta^{\mu})-S^{\mu\alpha\beta}\partial_{\mu}(\beta\omega_{\alpha\beta})-\beta\omega_{\alpha\beta}\partial_{\mu}S^{\mu\alpha\beta}+\partial_{\mu}Q^{\mu}\nonumber\\
& ~~~~~~~=\partial_{\mu}(p\beta^{\mu})+T^{\mu\nu}_{(0)}\partial_{\mu}\beta_{\nu}-S^{\mu\alpha\beta}_{(0)}\partial_{\mu}(\beta\omega_{\alpha\beta})+\left(\partial_{\mu}\beta_{\nu}+2\beta\omega_{\mu\nu}\right)T^{\mu\nu}_{(1a)}+T^{\mu\nu}_{(1s)}\partial_{\mu}\beta_{\nu}\nonumber\\
& ~~~~~~~~~~~~~~~~~~~~~~~~~~~~~~~~~~~~~ -S^{\mu\alpha\beta}_{(1)}\partial_{\mu}(\beta\omega_{\alpha\beta})+\partial_{\mu}Q^{\mu}. 
\label{equ59ver2}
\end{align}
To obtain the last line of the above equation we used the hydrodynamic equations~\eqref{eq1} and \eqref{eq2}. Moreover, using thermodynamic relations it can be easily shown that, 
\begin{align}
\partial_{\mu}(p\beta^{\mu})+T^{\mu\nu}_{(0)}\partial_{\mu}\beta_{\nu}-S^{\mu\alpha\beta}_{(0)}\partial_{\mu}(\beta\omega_{\alpha\beta})=0,
\end{align}
which, when used in Eq.~\eqref{equ59ver2}, leads to Eq.~\eqref{eq24}, i.e.,  
\begin{align}
\partial_{\mu}s^{\mu}_{\rm IS}= \left(\partial_{\mu}\beta_{\nu}+2\beta\omega_{\mu\nu}\right)T^{\mu\nu}_{(1a)}+T^{\mu\nu}_{(1s)}\partial_{\mu}\beta_{\nu} -S^{\mu\alpha\beta}_{(1)}\partial_{\mu}(\beta\omega_{\alpha\beta})+\partial_{\mu}Q^{\mu}. 
\end{align}

\section{Derivation of Eq.~\eqref{equ22ver2}}
\label{appendixC}
We start with Eq.~\eqref{eq24}, 
\begin{align}
\partial_{\mu}s^{\mu}_{\rm IS} & = \left(\partial_{\mu}\beta_{\nu}+2\beta\omega_{\mu\nu}\right)T^{\mu\nu}_{(1a)}+T^{\mu\nu}_{(1s)}\partial_{\mu}\beta_{\nu} -S^{\mu\alpha\beta}_{(1)}\partial_{\mu}(\beta\omega_{\alpha\beta})+\partial_{\mu}Q^{\mu}\nonumber\\
& = 2\beta\omega_{\mu\nu}T^{\mu\nu}_{(1a)}+T^{\mu\nu}_{(1a)}\partial_{\mu}\beta_{\nu}+T^{\mu\nu}_{(1s)}\partial_{\mu}\beta_{\nu}-S^{\mu\alpha\beta}_{(1)}\partial_{\mu}(\beta\omega_{\alpha\beta})+\partial_{\mu}Q^{\mu}.
\label{equ63ver2}
\end{align}
Using the explicit form of $T^{\mu\nu}_{(1s)}$ and $T^{\mu\nu}_{(1a)}$ it has been already shown in Ref.~\cite{Daher:2022xon} the first three terms in the above equation can be expressed as, 
\begin{align}
& 2\beta\omega_{\mu\nu}T^{\mu\nu}_{(1a)}+T^{\mu\nu}_{(1a)}\partial_{\mu}\beta_{\nu}+T^{\mu\nu}_{(1s)}\partial_{\mu}\beta_{\nu}\nonumber\\
=&-\beta h^{\mu}\big(\beta \nabla_{\mu}T-Du_{\mu}\big)+\beta \pi^{\mu\nu}\sigma_{\mu\nu}+\beta\Pi \theta\nonumber\\
    & -\beta q^{\mu}\left(\beta \nabla_{\mu}T+Du_{\mu}-4 \omega_{\mu\nu}u^{\nu}\right)\nonumber\\
     & +\phi^{\mu\nu}\left(\Omega_{\mu\nu}+2\beta \Delta^{\alpha}_{~\mu}\Delta^{\beta}_{~\nu}\omega_{\alpha\beta}\right) 
     \label{equ64ver2}
\end{align}
Here we defined $\Omega_{\mu\nu}\equiv\Delta^{\alpha}_{~\mu}\Delta^{\beta}_{~\nu}\partial_{[\alpha}\beta_{\beta]} = \beta \nabla_{[\mu} u_{\nu]}$. The tensor  $\sigma_{\mu\nu}=\nabla_{(\mu}u_{\nu)} -\frac{1}{3}\theta\Delta_{\mu\nu}$ is traceless, i.e. $\sigma^{\mu}_{~\mu}=0$, and  orthogonal to the fluid four velocity, i.e. $\sigma^{\mu\nu}u_{\mu}=0=\sigma^{\mu\nu}u_{\nu}$. 
Now let us consider the fourth term in Eq.~\eqref{equ63ver2},
\begin{align}
-S^{\mu\alpha\beta}_{(1)}\partial_{\mu}(\beta\omega_{\alpha\beta}) =&-\left(2u^{[\alpha}\Delta^{\mu\beta]}\Phi+2u^{[\alpha}\tau_{(s)}^{\mu\beta]}+2u^{[\alpha}\tau_{(a)}^{\mu\beta]}+\Theta^{\mu\alpha\beta}\right)\partial_{\mu}(\beta\omega_{\alpha\beta})\nonumber\\
 =&-2u^{[\alpha}\Delta^{\mu\beta]}\Phi \nabla_{\mu}(\beta\omega_{\alpha\beta})-2u^{[\alpha}\tau_{(s)}^{\mu\beta]}\nabla_{\mu}(\beta\omega_{\alpha\beta})-2u^{[\alpha}\tau_{(a)}^{\mu\beta]}\nabla_{\mu}(\beta\omega_{\alpha\beta})-\Theta^{\mu\alpha\beta}\nabla_{\mu}(\beta\omega_{\alpha\beta})\nonumber\\
 =&-2\Phi u^{\alpha}\nabla^{\beta}(\beta\omega_{\alpha\beta})-2u^{\alpha}\tau^{\mu\beta}_{(s)}\nabla_{\mu}(\beta\omega_{\alpha\beta})-2u^{\alpha}\tau^{\mu\beta}_{(a)}\nabla_{\mu}(\beta\omega_{\alpha\beta})-\Theta^{\mu\alpha\beta}\nabla_{\mu}(\beta\omega_{\alpha\beta})\nonumber\\
 =&-2\Phi u^{\alpha}\nabla^{\beta}(\beta\omega_{\alpha\beta})-\tau_{\mu\beta(s)}u^{\alpha}\left(\Delta^{\gamma\mu}\Delta^{\rho\beta}+\Delta^{\gamma\beta}\Delta^{\mu\rho}-\frac{2}{3}\Delta^{\gamma\rho}\Delta^{\mu\beta}\right)\nabla_{\gamma}(\beta\omega_{\alpha\rho})\nonumber\\
&-\tau_{\mu\beta(a)}u^{\alpha}\left(\Delta^{\gamma\mu}\Delta^{\beta\rho}-\Delta^{\mu\rho}\Delta^{\beta\gamma}\right)\nabla_{\gamma}(\beta\omega_{\alpha\rho})-\Theta_{\mu\alpha\beta}\Delta^{\alpha\delta}\Delta^{\beta\rho}\Delta^{\mu\gamma}\nabla_{\gamma}(\beta\omega_{\delta\rho})
\label{equ65ver2}
\end{align}
Using Eqs.~\eqref{equ64ver2} and \eqref{equ65ver2} in Eq.~\eqref{equ63ver2} we find,
\begin{align}
\partial_{\mu}s^{\mu}_{\rm IS}= &-\beta h^{\mu}\left(\beta \nabla_{\mu}T-Du_{\mu}\right)+\beta\pi^{\mu\nu}\sigma_{\mu\nu}+\beta\Pi \theta \nonumber\\
&-\beta q^{\mu}\left(\beta \nabla_{\mu}T+Du_{\mu}-4 \omega_{\mu\nu}u^{\nu}\right) +\phi^{\mu\nu}\left(\Omega_{\mu\nu}+2\beta \Delta^{\alpha}_{~\mu}\Delta^{\beta}_{~\nu}\omega_{\alpha\beta}\right)\nonumber\\
& -2\Phi u^{\alpha}\nabla^{\beta}(\beta\omega_{\alpha\beta})-\tau_{\mu\beta(s)}u^{\alpha}\left(\Delta^{\gamma\mu}\Delta^{\rho\beta}+\Delta^{\gamma\beta}\Delta^{\mu\rho}-\frac{2}{3}\Delta^{\gamma\rho}\Delta^{\mu\beta}\right)\nabla_{\gamma}(\beta\omega_{\alpha\rho})\nonumber\\
& -\tau_{\mu\beta(a)}u^{\alpha}\left(\Delta^{\gamma\mu}\Delta^{\beta\rho}-\Delta^{\mu\rho}\Delta^{\beta\gamma}\right)\nabla_{\gamma}(\beta\omega_{\alpha\rho})-\Theta_{\mu\alpha\beta}\Delta^{\alpha\delta}\Delta^{\beta\rho}\Delta^{\mu\gamma}\nabla_{\gamma}(\beta\omega_{\delta\rho})+\partial_{\mu}Q^{\mu}. 
\label{}
\end{align}

\section{Explicit expressions for $\mathcal{D}$, $\mathcal{A}^{\mu}$, $\mathcal{B}^{\mu}$, $\mathcal{C}^{\mu\nu}$, $\mathcal{E}^{\mu\nu}$, $\mathcal{F}$, $\mathcal{G}^{\mu\nu}$, $\mathcal{H}^{\mu\nu}$, and $\mathcal{I}^{\alpha\mu\nu}$ }
\label{appendixD}
The first step in deriving the following scalars, vectors, and tensors starts by taking the partial derivative of $Q^{\mu}$ in~Eq.~\eqref{eq19}. Note that the partial derivative of the parameters $a_{i}, \tilde{a}_{i}, b_{i}, \tilde{b}_{i},$ and $c_{i}$ is not zero. The next step is to collect all terms having common dissipative current. In such a process, one can encounter terms of two different dissipative currents, for example, `$\pi_{\mu\nu}h^{\nu}\nabla^{\mu}b_{2}$'. For that, we've introduced the constants $l$ and $\tilde{l}$ such that 
\begin{align}
    \pi_{\mu\nu}h^{\nu}\nabla^{\mu}b_{2}=l_{h\pi} \pi_{\mu\nu}h^{\nu}\nabla^{\mu}b_{2}+(1-l_{h\pi}) \pi_{\mu\nu}h^{\nu}\nabla^{\mu}b_{2}
\end{align}
Following the above procedure we obtain, 
\begin{align}
\mathcal{D}=&a_1\Pi\theta+\Pi Da_1+2a_1D\Pi+(1-l_{\Pi h})h^{\mu}\nabla_{\mu}b_1-b_1(1-\tilde{l}_{\Pi h})h^{\mu}Du_{\mu}+b_1\nabla_{\mu}h^{\mu}+l_{\Pi q}q^{\mu}\nabla_{\mu}b_4\nonumber\\
&-\tilde{l}_{\Pi q}b_4 q^{\mu}Du_{\mu}+b_4\nabla_{\mu}q^{\mu}+l_{\Theta\Pi}\Theta^{\alpha\mu\nu}\Delta_{\alpha\mu}\nabla_{\nu}c_3-\tilde{l}_{\Theta\Pi}c_3\Delta_{\alpha\mu}\Theta^{\alpha\mu\nu}Du_{\nu}+c_3\Delta_{\alpha\beta}\nabla_{\mu}\Theta^{\alpha\beta\mu}.
\label{equ78ver2}
\end{align}
\begin{align}
\mathcal{A}^{\mu}=& a_3h^{\mu}\theta+h^{\mu}Da_3+2a_3Dh^{\mu}+l_{\Pi h}\Pi\nabla^{\mu}b_1+b_1\nabla^{\mu}\Pi-b_1\tilde{l}_{\Pi h}\Pi Du^{\mu}+l_{\pi h}\pi^{\lambda\mu}\nabla_{\lambda}b_2+b_2\nabla_{\lambda}\pi^{\lambda\mu}\nonumber\\
&-b_2\tilde{l}_{\pi h}\pi^{\lambda\mu}Du_{\lambda}+l_{\phi h}\phi^{\lambda\mu}\nabla_{\lambda}b_3+b_3\nabla_{\lambda}\phi^{\lambda\mu}-b_3\tilde{l}_{\phi h}\phi^{\lambda\mu}Du_{\lambda}+l_{\Phi h}\Phi\nabla^{\mu}\tilde{b}_1+\tilde{b}_1\nabla^{\mu}\Phi-\tilde{b}_1\tilde{l}_{\Phi h}\Phi Du^{\mu}\nonumber\\
&+l_{\tau_s h}\tau^{\lambda\mu}_{(s)}\nabla_{\lambda}\tilde{b}_2+\tilde{b}_2\nabla_{\lambda}\tau^{\lambda\mu}_{(s)}-\tilde{b}_2\tilde{l}_{\tau_{s}h}\tau^{\lambda\mu}_{(s)}Du_{\lambda}+l_{\tau_a h}\tau^{\lambda\mu}_{(a)}\nabla_{\lambda}\tilde{b}_3+\tilde{b}_3\nabla_{\lambda}\tau^{\lambda\mu}_{(a)}-\tilde{b}_3\tilde{l}_{\tau_{a}h}\tau^{\lambda\mu}_{(a)}Du_{\lambda}.
\label{equ79ver2}
\end{align}
\begin{align}
\mathcal{B}^{\mu}=&a_4q^{\mu}\theta+q^{\mu}Da_4+2a_4Dq^{\mu}+(1-l_{\Pi q})\Pi\nabla^{\mu}b_4+b_4\nabla^{\mu}\Pi-b_4(1-\tilde{l}_{\Pi q})\Pi Du^{\mu}+(1-l_{\pi q})\pi^{\lambda\mu}\nabla_{\lambda}b_5
\nonumber\\
&+b_5\nabla_{\lambda}\pi^{\lambda\mu}-b_5(1-\tilde{l}_{\pi q})\pi^{\lambda\mu}Du_{\lambda}+l_{\phi q}\phi^{\lambda\mu}\nabla_{\lambda}b_6+b_6\nabla_{\lambda}\phi^{\lambda\mu}-b_6\tilde{l}_{\phi q}\phi^{\lambda\mu}Du_{\lambda}+l_{\Phi q}\Phi\nabla^{\mu}\tilde{b}_4+\tilde{b}_4\nabla^{\mu}\Phi
\nonumber\\
&-\tilde{b}_4\tilde{l}_{\Phi q}\Phi Du^{\mu}+l_{\tau_s q}\tau^{\lambda\mu}_{(s)}\nabla_{\lambda}\tilde{b}_5+\tilde{b}_5\nabla_{\lambda}\tau^{\lambda\mu}_{(s)}-\tilde{b}_5\tilde{l}_{\tau_{s}q}\tau^{\lambda\mu}_{(s)}Du_{\lambda}+l_{\tau_a q}\tau^{\lambda\mu}_{(a)}\nabla_{\lambda}\tilde{b}_6+\tilde{b}_6\nabla_{\lambda}\tau^{\lambda\mu}_{(a)}-\tilde{b}_6\tilde{l}_{\tau_{a}q}\tau^{\lambda\mu}_{(a)}Du_{\lambda}.
\label{equ80ver2}
\end{align}
\begin{align}
\mathcal{C}^{\mu\nu} =&a_2\theta \pi^{\mu\nu}+\pi^{\mu\nu}Da_2+2a_2 D\pi^{\mu\nu}+(1-l_{\pi h})h^{(\nu}\nabla^{\mu)}b_2-b_2(1-\tilde{l}_{\pi h})h^{(\nu}Du^{\mu)}+b_2\nabla^{(\mu}h^{\nu)}\nonumber\\
&+l_{\pi q}q^{(\nu}\nabla^{\mu)}b_5-\tilde{l}_{\pi q}b_5 q^{(\nu}Du^{\mu)}+b_5\nabla^{(\mu}q^{\nu)}+l_{\Theta\pi}\Theta^{(\mu\nu)\alpha}\nabla_{\alpha}c_4-\tilde{l}_{\Theta\pi}c_4\Theta^{(\mu\nu)\alpha}Du_{\alpha}+c_4\nabla_{\alpha}\Theta^{(\mu\nu)\alpha}~.
\label{equ81ver2}
\end{align}
\begin{align}
\mathcal{E}^{\mu\nu}=&a_5\theta\phi^{\mu\nu}+\phi^{\mu\nu}Da_5+2a_5 D\phi^{\mu\nu}+(1-l_{\phi h})h^{[\nu}\nabla^{\mu]}b_3-b_3(1-\tilde{l}_{\phi h})h^{[\nu}Du^{\mu]}+b_3\nabla^{[\mu}h^{\nu]}\nonumber\\
&+(1-l_{\phi q})q^{[\nu}\nabla^{\mu]}b_6-b_6(1-\tilde{l}_{\phi q})q^{[\nu}Du^{\mu]}+b_6\nabla^{[\mu}q^{\nu]}+l_{\Theta\phi}\Theta^{\lambda\mu\nu}\nabla_{\lambda}c_1-\tilde{l}_{\Theta\phi}c_1\Theta^{\lambda\mu\nu}Du_{\lambda}
\nonumber\\
&+c_3\nabla_{\lambda}\Theta^{\lambda\mu\nu}+k_{\Theta\phi}\Theta^{[\mu\nu]\lambda}\nabla_{\lambda}c_7-\tilde{k}_{\Theta\phi}c_7\Theta^{[\mu\nu]\lambda}Du_{\lambda}+c_7\nabla_{\lambda}\Theta^{[\mu\nu]\lambda}~.
\label{equ82ver2}
\end{align}
\begin{align}
\mathcal{F}=&\tilde{a}_1\theta\Phi+\Phi D\tilde{a}_1+2\tilde{a}_1D\Phi+(1-l_{\Phi h})h^{\mu}\nabla_{\mu}\tilde{b}_1-(1-\tilde{l}_{\Phi h})\tilde{b}_1h^{\mu}Du_{\mu}+\tilde{b}_1\nabla_{\mu}h^{\mu}+(1-l_{\Phi q})q^{\mu}\nabla_{\mu}\tilde{b}_4\nonumber\\
&-(1-\tilde{l}_{\Phi q})\tilde{b}_4 q^{\mu}Du_{\mu}+\tilde{b}_4\nabla_{\mu}q^{\mu}+l_{\Theta\Phi}\Theta^{\alpha\mu\nu}\Delta_{\alpha\mu}\nabla_{\nu}c_5-\tilde{l}_{\Theta\Pi}c_5\Delta_{\alpha\mu}\Theta^{\alpha\mu\nu}Du_{\nu}+c_5\Delta_{\alpha\beta}\nabla_{\mu}\Theta^{\alpha\beta\mu}.
\label{equ83ver2}
\end{align}
\begin{align}
\mathcal{G}^{\mu\nu} =&\tilde{a}_2\theta \tau^{\mu\nu}_{(s)}+\tau^{\mu\nu}_{(s)}D\tilde{a}_2+2\tilde{a}_2 D\tau^{\mu\nu}_{(s)}+(1-l_{\tau_{s} h})h^{(\nu}\nabla^{\mu)}\tilde{b}_2-\tilde{b}_2(1-\tilde{l}_{\tau_s h})h^{(\nu}Du^{\mu)}+\tilde{b}_2\nabla^{(\mu}h^{\nu)}\nonumber\\
&+(1-l_{\tau_s q})q^{(\nu}\nabla^{\mu)}\tilde{b}_5-(1-\tilde{l}_{\tau_s q})\tilde{b}_5 q^{(\nu}Du^{\mu)}+\tilde{b}_5\nabla^{(\mu}q^{\nu)}+l_{\Theta\tau_s}\Theta^{(\mu\nu)\lambda}\nabla_{\lambda}c_6-\tilde{l}_{\Theta\tau_s}c_6\Theta^{(\mu\nu)\lambda}Du_{\lambda}\nonumber\\
&+c_6\nabla_{\lambda}\Theta^{(\mu\nu)\lambda}~.
\label{equ84ver2}
\end{align}
\begin{align}
\mathcal{H}^{\mu\nu} =&\tilde{a}_3\theta\tau^{\mu\nu}_{(a)}+\tau^{\mu\nu}_{(a)}D\tilde{a}_3+2\tilde{a}_3 D\tau^{\mu\nu}_{(a)}+(1-l_{\tau_a h})h^{[\nu}\nabla^{\mu]}\tilde{b}_3-\tilde{b}_3(1-\tilde{l}_{\tau_a h})h^{[\nu}Du^{\mu]}+\tilde{b}_3\nabla^{[\mu}h^{\nu]}\nonumber\\
&+(1-l_{\tau_a q})q^{[\nu}\nabla^{\mu]}\tilde{b}_6-\tilde{b}_6(1-\tilde{l}_{\tau_a q})q^{[\nu}Du^{\mu]}+\tilde{b}_6\nabla^{[\mu}q^{\nu]}+l_{\Theta\tau_a}\Theta^{\lambda\mu\nu}\nabla_{\lambda}c_2-\tilde{l}_{\Theta\tau_a}c_2\Theta^{\lambda\mu\nu}Du_{\lambda}
\nonumber\\
&+c_2\nabla_{\lambda}\Theta^{\lambda\mu\nu}+k_{\Theta\tau_a}\Theta^{[\mu\nu]\lambda}\nabla_{\lambda}c_8-\tilde{k}_{\Theta\tau_a}c_8\Theta^{[\mu\nu]\lambda}Du_{\lambda}+c_8\nabla_{\lambda}\Theta^{[\mu\nu]\lambda}~.
\label{equ85ver2}
\end{align}
\begin{align}
\mathcal{I}^{\alpha\mu\nu}  =&\tilde{a}_4\theta\Theta^{\alpha\mu\nu}+ \Theta^{\alpha\mu\nu}D\tilde{a}_4+2\tilde{a}_4D\Theta^{\alpha\mu\nu} + (1-l_{\Theta \phi})\phi^{\mu\nu}\nabla^{\alpha}c_1-(1-\tilde{l}_{\Theta \phi})c_1\phi^{\mu\nu}Du^{\alpha}+c_1\nabla^{\alpha}\phi^{\mu\nu}\nonumber\\
&+ (1-l_{\Theta \tau_a})\tau_{(a)}^{\mu\nu}\nabla^{\alpha}c_2-(1-\tilde{l}_{\Theta \tau_a})c_2\tau^{\mu\nu}_{(a)}Du^{\alpha}+c_2\nabla^{\alpha}\tau^{\mu\nu}_{(a)}+(1-l_{\Theta\Pi})\Pi\Delta^{\alpha[\mu}\nabla^{\nu]}c_3\nonumber\\
&-(1-\tilde{l}_{\Theta\Pi})c_3\Pi\Delta^{\alpha[\mu}Du^{\nu]}+c_3\Delta^{\alpha[\mu}\nabla^{\nu]}\Pi+(1-l_{\Theta\Phi})\Phi\Delta^{\alpha[\mu}\nabla^{\nu]}c_5-(1-\tilde{l}_{\Theta\Phi})c_5\Phi\Delta^{\alpha[\mu}Du^{\nu]}
\nonumber\\
&+c_5\Delta^{\alpha[\mu}\nabla^{\nu]}\Phi+(1-l_{\Theta\pi})\pi^{\alpha[\mu}\nabla^{\nu]}c_4-(1-\tilde{l}_{\Theta\pi})c_4\pi^{\alpha[\mu}Du^{\nu]}+c_4\nabla^{[\nu}\pi^{\alpha\mu]}+(1-l_{\Theta\tau_s})\tau_{(s)}^{\alpha[\mu}\nabla^{\nu]}c_6
\nonumber\\
&-(1-\tilde{l}_{\Theta\tau_s})c_6\tau_{(s)}^{\alpha[\mu}Du^{\nu]}+c_6\nabla^{[\nu}\tau_{(s)}^{\alpha\mu]} +(1-k_{\Theta\phi})\phi^{\alpha[\mu}\nabla^{\nu]}c_7-(1-\tilde{k}_{\Theta\phi})c_7\phi^{\alpha[\mu}Du^{\nu]}+c_7\nabla^{[\nu}\phi^{\alpha\mu]}\nonumber\\
&+(1-k_{\Theta\tau_a})\tau_{(a)}^{\alpha[\mu}\nabla^{\nu]}c_8-(1-\tilde{k}_{\Theta\tau_a})c_8\tau_{(a)}^{\alpha[\mu}Du^{\nu]}+c_8\nabla^{[\nu}\tau_{(a)}^{\alpha\mu]}~.
\label{equ86ver2}
\end{align}

\bibliography{ref.bib}{}
\bibliographystyle{utphys}
%\end{thebibliography}
\end{document}